\documentclass[letterpaper,twocolumn,10pt]{article}
\usepackage{usenix2019,epsfig,endnotes}

\usepackage[T1]{fontenc}
\usepackage[utf8]{inputenc}
\usepackage{letltxmacro}
\usepackage{todonotes}
\usepackage{booktabs}
\usepackage{amssymb}
\usepackage{xspace}
\usepackage[hyphens]{url}
\usepackage{hyperref}
\usepackage{listings}
\usepackage{paralist}
\usepackage{pifont}
\usepackage{enumitem}
\usepackage{multirow}
\usepackage{pbox}
\usepackage[detect-all]{siunitx}
\usepackage{enumitem}
\usepackage{float}
\usepackage{subcaption}

\lstdefinelanguage
[aarch64]{Assembler}
[x86masm]{Assembler}
{
    morekeywords={mov,autiasp,paciasp,retaa,reta,paciasp,autib,pacib,pacia,bla,and,stp,ldp,autda,pacda,
    b.ge,autia,b,movk,Xd,blraa,eor,Xptr,Xmod,
    x29,x30,sp,str,ldr,stp,ldp,x0,lr,fp,x,p,d,c}
}

\lstdefinestyle{customasm}{
  captionpos=b,
  frame=single,
  language=[aarch64]Assembler,
  commentstyle=\itshape\color{purple!40!black},
}
\newif\ifabridged
\newif\ifnotabridged
\newif\ifanonymous
\newif\ifnotanonymous

\ifdefined\isabridged
\abridgedtrue
\fi

\ifabridged\notabridgedfalse
\else\notabridgedtrue
\fi

\ifdefined\isanonymous
\anonymoustrue
\fi

\ifanonymous\notanonymousfalse
\else\notanonymoustrue
\fi
\LetLtxMacro{\todonote}{\todo}
\renewcommand{\todo}[2][]
{\todonote[inline, caption={#2}, size=\footnotesize, #1]
{\renewcommand{\baselinestretch}{0.5}\selectfont#2\par}}

\newcommand{\SHORTNAME}{\protect{PARTS}\xspace}
\newcommand{\PA}{\protect{PA}\xspace}
\newcommand{\ARMPA}{\protect{PA}\xspace}
\newcommand{\ARMPAFULL}{\protect{ARMv8-A PA}\xspace}
\newcommand{\SHORTNAMELIB}{\protect{PARTSlib}\xspace}

\newcommand{\LONGNAME}{\protect{Pointer Authentication Run-Time Safety}\xspace}

\newcommand{\instr}[1]{\texttt{\lowercase{#1}}}

\newcommand{\dOpt}{\ding{182}\xspace}
\newcommand{\dAArch}{\ding{183}\xspace}
\newcommand{\dBackend}{\ding{184}\xspace}
\newcommand{\dLib}{\ding{185}\xspace}

\newcommand{\dOne}{\ding{182}\xspace}
\newcommand{\dTwo}{\ding{183}\xspace}
\newcommand{\dThree}{\ding{184}\xspace}
\newcommand{\dFour}{\ding{185}\xspace}

\newcommand{\dCOne}{\ding{192}\xspace}
\newcommand{\dCTwo}{\ding{193}\xspace}
\newcommand{\dCThree}{\ding{194}\xspace}
\newcommand{\dCFour}{\ding{195}\xspace}
\newcommand{\dCFive}{\ding{196}\xspace}
\newcommand{\dCSix}{\ding{197}\xspace}

\newcommand{\tableNo}{\ding{55}}
\newcommand{\tableYes}{\ding{51}}

\graphicspath{{figures/}}
\setlist[enumerate]{topsep=0pt,itemsep=-1ex,partopsep=1ex,parsep=1ex}
\setlist[itemize]{topsep=0pt,itemsep=-1ex,partopsep=1ex,parsep=1ex}

\title{\Large\bf{}PAC it up: Towards Pointer Integrity using ARM Pointer Authentication
\ifabridged{}
\footnotemark{}
\fi{}}
\ifnotanonymous
\author{
  {\rm Hans Liljestrand}\\
  {\small Aalto University, Finland}\\
  {\small Huawei Technologies Oy, Finland}\\
  {\small\tt hans.liljestrand@aalto.fi}
  \\\\
  {\rm Carlos Chinea Perez}\\
  {\small Huawei Technologies Oy, Finland}\\
  {\small\tt carlos.chinea.perez@huawei.com}
  \and
  {\rm Thomas Nyman}\\
  {\small Aalto University, Finland}\\
  {\small\tt thomas.nyman@aalto.fi}
  \\\\\\
  {\rm Jan-Erik Ekberg}\\
  {\small Huawei Technologies Oy, Finland}\\
    {\small Aalto University, Finland}\\
  {\small\tt jan.erik.ekberg@huawei.com}
  \and
  {\rm Kui Wang}\\
  {\small Huawei Technologies Oy, Finland}\\
  {\small Tampere University of Technology, Finland}\\
  {\small\tt wang.kui1@huawei.com}
  \\\\
  {\rm N. Asokan}\\
  {\small Aalto University, Finland}\\
  {\small\tt asokan@acm.org}
}
\fi

\makeatletter
\def\blfootnote{\xdef\@thefnmark{}\@footnotetext}
\makeatother

\begin{document}
\date{}
\maketitle

\ifabridged{}
\blfootnote{Extended version of this article is available as a technical report~\cite{Liljestrand19}.}
\fi{}
\begin{abstract}

Run-time attacks against programs written in memory-unsafe programming languages (e.g., C and C++) remain a prominent threat against computer systems.
The prevalence of techniques like return-oriented programming (ROP) in attacking real-world systems has prompted major processor manufacturers to design hardware-based countermeasures against specific classes of run-time attacks.
An example is the recently added support for \emph{pointer authentication} (\ARMPA) in the ARMv8-A processor architecture, commonly used in devices like smartphones.
PA is a low-cost technique to authenticate pointers so as to resist memory vulnerabilities.
It has been shown to enable practical protection against memory vulnerabilities that corrupt return addresses or function pointers.
However, so far, \ARMPA has received very little attention as a general purpose protection mechanism to harden software against various classes of memory attacks.

In this paper, we use \ARMPA{} to build novel defenses against various classes of run-time attacks, including the first \PA-based mechanism for data pointer integrity.
We present \SHORTNAME, an instrumentation framework that integrates our \PA-based defenses into the LLVM compiler and the GNU/Linux operating system and show, via systematic evaluation, that \SHORTNAME provides better protection than current solutions at a reasonable performance overhead.

\end{abstract}
\section{Introduction}
\label{sec:introduction}

Memory corruption vulnerabilities, such as buffer overflows, continue to be a prominent threat against modern software applications written in memory-unsafe programming languages, like C and C++. Theses vulnerabilities can be exploited to overwrite data in program memory. By overwriting control data, such as code pointers and return addresses, attackers can redirect execution to attacker-chosen locations.
\emph{Return-oriented programming} (ROP)~\cite{Shacham07} is a well known technique that allows the attacker to leverage corrupted control-data and pre-existing code sequences to construct powerful (Turing-complete) attacks without the need to inject code into the victim program.
By overwriting non-control data, such as variables used for decision making, attackers can also influence program behavior without breaking the program's \emph{control-flow integrity} (CFI)~\cite{Abadi09}.
Such attacks can cause the program to leak sensitive data or escalate attacker privileges.
Recent work has shown that non-control-data attacks can also be generalized to achieve Turing-completeness. Such \emph{data-oriented programming} (DOP) attacks~\cite{Hu16} are difficult to defend against, and are an appealing attack technique for future run-time exploitation.
Software defenses against run-time attacks can offer strong security guarantees, but their usefulness is limited by high performance overhead, or requiring significant changes to system software architecture.
Consequently, deployed solutions (e.g., Microsoft EMET~\cite{MS-EMET}) trade off security for performance.
Various hardware-assisted defenses in the research literature~\cite{Devietti08,Woodruff14,Watson15,Davi15,Song16,Tsampas17,Nyman17b,Roessler18} can drastically improve the efficiency  of attack detection, but the majority of such defenses are unlikely to ever be deployed as they require invasive changes to the underlying processor architecture.
However, the prevalence of advanced attack techniques (e.g, ROP) in modern run-time exploitation has prompted major processor vendors to integrate security primitives into their processor designs to thwart specific attacks efficiently~\cite{Intel-CET,Oleksenko17,Qualcomm17}.
Recent additions to the ARMv8-A architecture~\cite{ARMv8A} include new instructions for \emph{pointer authentication} (\PA).
PA uses cryptographic message authentication codes (MACs), referred to as \emph{pointer authentication codes} (PACs), to protect the integrity of pointers.
However, \PA is vulnerable to \emph{pointer reuse} attacks where an authenticated pointer is substituted with another~\cite{Qualcomm17}.
Practical \ARMPA{}-based defenses must minimize the scope of such substitution.

\paragraph*{Goals and Contributions} In this work, we further the security analysis of \ARMPAFULL{} by categorizing pointer reuse attacks, and show that \PA enables practical defenses against several classes of run-time attacks.
We propose an enhanced scheme for \emph{pointer signing} that enforces \emph{pointer integrity} for all code and data pointers. We also propose \emph{run-time type safety} which constrains pointer substitution attacks by ensuring the pointer is of the correct type. Pointer signing and run-time type safety are effective against both control-flow and data-oriented attacks. Finally, we design and implement \emph{Pointer Authentication Run-Time Safety} (\SHORTNAME), a compiler instrumentation framework that leverages \PA{} to realize our proposed defenses.
We evaluate the security and practicality of \SHORTNAME{}
to demonstrate its effectiveness against memory corruption attacks.
Our main contributions are:
\begin{itemize}
  \item \emph{Analysis:} A categorization and analysis of pointer reuse and other attacks against ARMv8-A pointer authentication (Section~\ref{sec:attacks-on-pa}).
  \item \emph{Design:} A scheme for using \emph{pointer integrity} to systematically defend against control-flow and data-oriented attacks, and \emph{run-time type safety}, a scheme for guaranteeing safety for data and code pointers at run-time (Section~\ref{sec:design}).
  \item \emph{Implementation:} \SHORTNAME{}, a compiler instrumentation framework that uses \PA to realize data pointer, code pointer, and return address signing (Section~\ref{sec:implementation}).
  \item \emph{Evaluation:}
    Systematic analysis of \SHORTNAME{} 
    showing that it has a reasonable performance overhead ($<0.5$\% average overhead for code-pointer and return address signing, $19.5$\% average overhead for data-pointer signing in nbench-byte (Section~\ref{sec:evaluation})) and provides better security guarantees than fully-precise static CFI
    (\ref{sec:discussion}).
\end{itemize}

\ifabridged{}
We make the source code of \SHORTNAME{} publicly available at \url{https://github.com/pointer-authentication}.
\fi{}

\section{Background}
\label{sec:background}

\subsection{Run-time attacks}
\label{sec:run-time-attacks}

Programs written in memory-unsafe languages are prone to memory errors like buffer-overflows, use-after-free errors and format string vulnerabilities~\cite{Szekeres2013}.
Traditional approaches for exploiting such errors by corrupting program code have been rendered largely ineffective by the widespread deployment of measures like data execution prevention (DEP). This has given rise to two new attack classes: \emph{control-flow attacks} and \emph{data-oriented attacks}~\cite{Chen05}.

\subsubsection{Control-flow attacks (on ARM)}
\label{sec:run-time-attacks-ARM}

Control-flow attacks exploit memory errors to hijack program execution by overwriting code pointers (function return addresses or function pointers).
Corrupting a code pointer can cause a control-flow transfer to anywhere in executable memory.
Corrupting the return address of a function can be used for ROP attacks, which are feasible on several architectures, including ARM~\cite{Kornau2009}.

ARM processors, similar to other RISC processor designs, have a dedicated Link Register (LR) that stores the return address.
LR is typically set during a function call by the Branch with Link (\instr{bl}) instruction.
An attacker cannot directly influence the value of LR, as it is unlikely for a program to contain instructions for directly modifying it.
However, nested function calls require the return address of a function to be stored on the stack before the next function call replaces the LR value.
While the return address is stored on the stack, an attacker can use a memory error to modify it to subsequently redirect the control flow on function return.
On both x86 and ARM, it is possible to perform ROP attacks without the use of return instructions.
Such attacks are collectively referred to as \emph{jump-oriented programming} (JOP)~\cite{Checkoway2010}.

Control-flow integrity (CFI)~\cite{Abadi09} is a prominent defense technique against control-flow attacks.
The  goal of CFI is to allow all the control flows present in a program's control-flow graph (CFG), while rejecting other flows.
\ifabridged{}
Widely deployed CFI solutions are less precise than state-of-the-art solutions presented in scientific literature.
\else
Practical deployment of CFI solutions must trade off precision with performance overhead.
Thus, widely deployed CFI solutions are less precise than state-of-the-art solutions presented in scientific literature.
\fi
\subsubsection{Data-oriented attacks}
\label{sec:run-time-attacks-DOP}

In contrast to control-flow attacks, \emph{data-oriented attacks} can influence program behavior without the need to modify code pointers.
Instead, they corrupt variables that influence the program's decision making, or leak sensitive information from program memory. Such attacks are called \emph{non-control-data attacks}.
Chen et al~\cite{Chen05} demonstrated a variety of non-control-data attacks for forging user credentials, changing security critical configuration parameters, bypassing security checks, and escalating privileges.
Recent work on DOP~\cite{Hu16} showed that non-control-data corruption can also enable expressive attacks
without compromising control-flow integrity.
DOP may compromise the input of individual program operations and chain together a chosen sequence of operations to achieve the intended functionality.

A data-oriented attack can in principle corrupt arbitrary program objects, but corrupting data pointers is often the preferred attack vector~\cite{Cowan03}.
In Chen et al.'s attack against the GHTTPD web server~\cite{Chen05}, a stack buffer overflow is used to corrupt a data pointer used in input string validation in order to bypass security checks on the input under the attacker's control.
Data pointers are also routinely corrupted in heap exploitation.
For instance, the ``\emph{House of Spirit}'' attack on Glibc\footnote{Team Shellphish repository of educational heap exploitation techniques: \url{https://github.com/shellphish/how2heap}}, involves corrupting a pointer returned by \texttt{malloc()} to trick subsequent \texttt{malloc()} calls into returning attacker controlled memory chunks.
The DOP attacks in~\cite{Hu16} also involve the corruption of pointers as a means to control which data is processed by vulnerable code.

\subsection{ARM Pointer Authentication}
\label{sec:pa}

\begin{figure}[tp]
\centering
\includegraphics[width=\columnwidth]{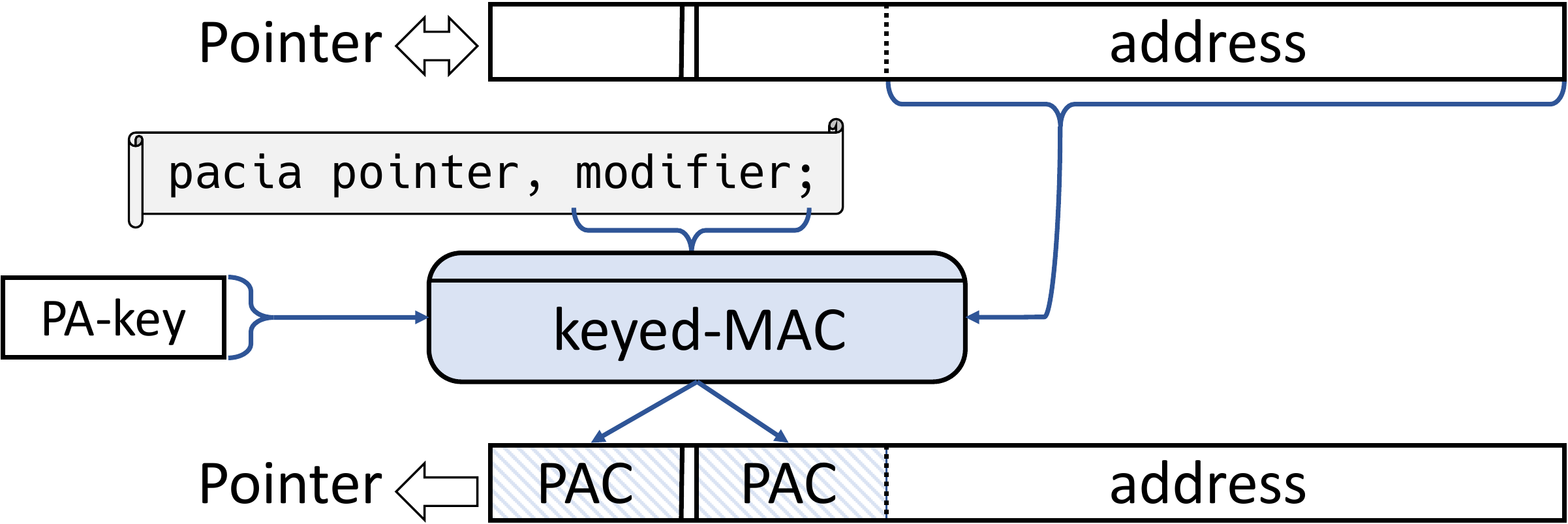}
\caption{
The PAC is created using key-specific PA instructions (\texttt{pacia}) and is a keyed MAC calculated over the pointer address and a modifier.
}
\label{fig:pa-construction}
\end{figure}

ARMv8.3-A includes a new feature called \emph{pointer authentication} (\PA).
\PA{} is intended for checking the integrity of pointers with minimal size and performance impact.
It is available when the processor executes in 64-bit ARM state (AArch64).
\PA{} adds instructions for creating and authenticating \emph{pointer authentication codes} (PACs).
The PAC is a tweakable message authentication code (MAC) calculated over the pointer value and a 64-bit \emph{modifier} as the tweak (Figure~\ref{fig:pa-construction}).
Different combinations of key and modifier pairs allow domain separation among different classes of authenticated pointers.
This prevents authenticated pointer values from being arbitrarily interchangeable with one another.
\ifnotabridged{}
Preventing, for example, attacks from using a function pointer as a return address, or vice versa.
\fi

The idea of using of MACs to protect pointers at run-time is not new. \emph{Cryptographic CFI} (CCFI)~\cite{Mashtizadeh15} uses MACs to protect control-flow data such as return addresses, function pointers, and vtable pointers. Unlike \ARMPAFULL, CCFI uses hardware-accelerated AES for speeding up MAC calculation. Run-time software checks are needed to compare the calculated MAC to a reference value. \PA, on the other hand, uses either QARMA~\cite{Avanzi17} or a manufacturer-specific MAC, and performs the MAC comparison in hardware.

\begin{figure}[tp]
\centering
\includegraphics[width=\columnwidth]{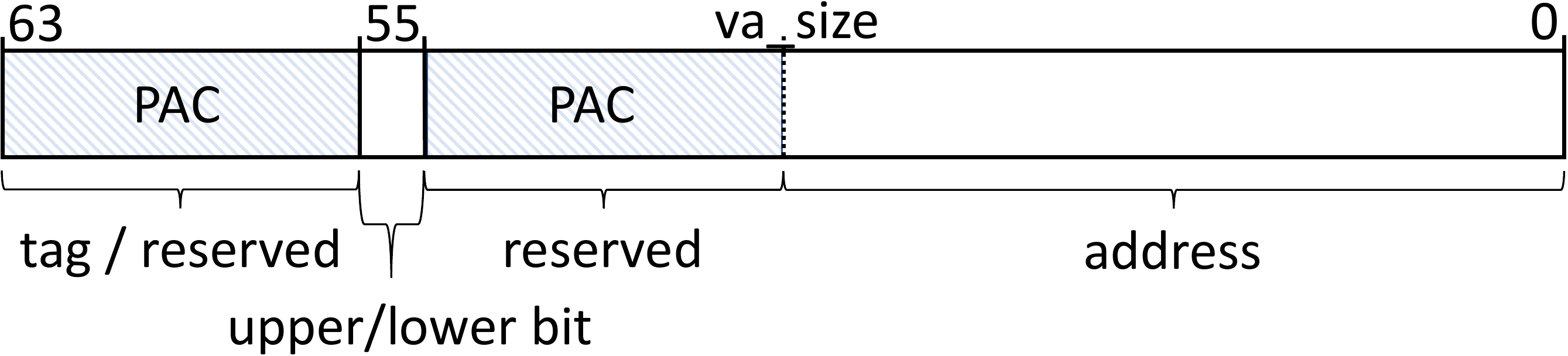}
\caption{
Pointer layout on 64-bit ARM\@. The PAC is stored in the reserved bits, and its size depends on the used virtual address range. If pointer tagging is disabled, then the PAC can also extend to the tag bits.
}
\label{fig:pointer_layout}
\end{figure}

64-bit ARM processors only use part of the 64-bit address space for virtual addresses (Figure~\ref{fig:pointer_layout}).
The PAC is stored in the remaining unused bits of the pointer.
On a default AArch64 Linux kernel configuration with 39 bit addresses and without address tagging~\cite[D4.1.4]{ARMv8A}, the PAC size is 24 bits.
However, depending on the memory addressing scheme and whether address tagging is used, the size of the PAC is between 3 and 31 bits~\cite{Qualcomm17}.
Security implications of the PAC size are discussed in Section~\ref{sec:discussion}.

\PA provides five different keys for PAC generation: two for code pointers, two for data pointers, and one for generic use.
The keys are stored in hardware registers configured to be accessible only from a higher privilege level: e.g., the kernel maintains the keys for a user space process, generating keys for each process at process \texttt{exec}.
The keys remain constant throughout the process lifetime, whereas the modifier is given in an instruction-specific register operand on each PAC creation and authentication (i.e., MAC verification).
Thus it can be used to describe the run-time context in which the pointer is created and used.
The modifier value is not necessarily confidential (see Section~\ref{sec:adversary-model}) but ideally such that it
\begin{inparaenum}[1)]
\item precisely describes the context of use in which the pointer is valid, and
\item cannot be influenced by the attacker.
\end{inparaenum}

\PA{} is used by instrumenting code with PAC creation and authentication instructions.
\PA{} instruction mnemonics are generally prefixed either with \instr{PAC} or \instr{AUT} for creation and authentication, respectively, followed by two characters that select one of the data or code keys.
For instance, the \instr{pacia} instruction in Figure~\ref{fig:pa-construction} will generate an authenticated pointer (\instr{pac}) based on the instruction (\instr{I}) A-key (\instr{A}).
Table~\ref{tab:pa-instructions} in Appendix~\ref{sec:appendix-instructions} provides a list of \PA{} instructions \ifabridged{} referred to in this paper\fi.
An authenticated pointer cannot be used directly, as the PAC embedded in the pointer value intentionally interferes with address translation.
The corresponding \PA{} authentication instruction (in this case, \instr{AUTIA}) removes the PAC from the pointer if authentication is successful, i.e., if the current pointer value, key and modifier for \instr{AUTIA} yields a PAC that matches the PAC embedded in the pointer. 
If authentication fails, the pointer is invalidated such that a dereference or call using the pointer will cause a memory translation fault.
Dedicated \PA{} instructions are encoded in NOP space; older processors without \PA{} support will ignore them.
\ifnotabridged{}
For code pointers, ARM has combined \PA{} instructions that can do authentication and branching in one instruction, but these are not backwards compatible.
For instance, the \instr{blra} (Branch with Link to Register, with pointer Authentication) instruction can be used to implement an indirect function call using an authenticated pointer.
\fi

\paragraph{Return address signing.}
\label{sec:back-qualcomm}

Qualcomm's return address signing scheme~\cite{Qualcomm17} is the first to make use of \ARMPAFULL.
It was first introduced in Linaro's GCC toolchain, but has been supported by mainline GCC since version 7.0\footnote{GCC return address signing and PA support is based on patches provided by ARM, \url{https://github.com/gcc-mirror/gcc/commit/06f29de13f48f7da8a8c616108f4e14a1d19b2c8}}.
It thwarts attacks that manipulate function return addresses through stack corruption (see Section~\ref{sec:run-time-attacks-ARM}) by ensuring that the return address in LR always contains a PAC when written to or retrieved from memory.
Listing~\ref{lst:backward-edge-cfi} shows an example.

The instrumentation adds \instr{paciasp} (\dCOne)\xspace at beginning of the function prologue, before the LR value is stored on the stack.
\instr{paciasp} adds a PAC tag using the current Stack Pointer (SP) value as the modifier.
Before function return, \instr{autiasp} (\dCTwo)\xspace authenticates the pointer and either removes the PAC or invalidates the pointer.
An alternative is to use the combined \instr{autiasp}+\instr{ret} instruction, \instr{retaa}, but it is not backwards-compatible with older processors.

\begin{lstlisting}[float,floatplacement=tp,style=customasm,
label={lst:backward-edge-cfi},
caption={
Return address signing using PA.
At funtion entry, \instr{paciasp} is used to create a PAC in LR (\dCOne).
The value is then authenticated with \instr{autiasp} before return (\dCTwo).
}]
function:
  paciasp              ; (*@\dCOne@*)create PAC
  stp FP, LR, [SP, #0] ; store LR
  ; ...
  ldp FP, LR, [SP, #0] ; load LR
  autiasp              ; (*@\dCTwo@*)authenticate
  ret                  ; return
\end{lstlisting}
The PAC cryptographically binds the return address to the current SP value.
It is valid only when authenticated using the same SP value as on PAC creation.
The goal is to limit the validity of the PAC to the function invocation that created it, thus preventing reuse of authenticated return addresses.

\section{Attacks on Pointer Authentication}
\label{sec:attacks-on-pa}

\PA prevents an attacker from injecting or forging pointer values.
This effectively prevents any attack that relies on corrupting pointers, resisting even attackers with \emph{arbitrary access to program memory}.

\ifnotabridged{}
To protect authenticated pointers, PA relies on the confidentiality of process-specific \PA{} keys and the immutability (but not confidentiality) of \PA{} modifier values.
\PA keys are managed by the kernel and never revealed to user space.
Although the keys are used by PAC creation and authentication operations in user space, such operations take place using dedicated \PA instructions, and direct access to the \PA key registers is subject to hardware-enforced access controls.
Consequently, our adversary model (Section~\ref{sec:adversary-model}) assumes that the attacker cannot read or modify the \PA keys.
\fi

The modifier value used in computing a PAC can depend on both static (e.g., a hard-coded value) and dynamic (e.g., the SP) information.
We assume that the program code itself is not confidential and that the attacker can learn how dynamic modifiers are generated and may infer their values.

\PA also relies on the security of the underlying cryptographic primitives.
In particular, an attacker may attempt to brute-force either the \PA keys themselves, or individual PAC values.
Sophisticated adversaries may even attempt cryptanalysis attacks based on known PAC values, or side-channels attacks against the hardware circuitry for computing PACs.
The security of the QARMA block cipher has already been analyzed~\cite{Zong16,Li18}. We leave the scrutiny of the cryptographic building blocks outside the scope of this paper.
Nevertheless, the limited PAC size means that guessing attacks are a potential concern.
We discuss the feasibility of brute-forcing PACs in Section~\ref{sec:eval-pac-entropy}.
Assuming proper precautions for the lifetime of \PA keys (see Section~\ref{sec:pa}), we do not consider guessing attacks the primary attack vector against PA\@.
However, the following concerns for the security of PA-based defenses remain:
\begin{inparaenum}[1)]
\item\label{item:pa-attack-generate} an attacker controlling the creation of PAC values, or
\item\label{item:pa-attack-reuse} an attacker \emph{reusing} previously authenticated pointers.
\end{inparaenum}

\paragraph{Malicious PAC generation.}

Attackers can potentially control PAC values in three ways, by controlling:
\begin{enumerate}
  \item \emph{the unauthenticated pointer value before PAC creation}: get an arbitrary authenticated pointer for any context with the same modifier and \PA key.
  \item \emph{control the PA modifier value}: get an authenticated pointer for a context with the same \PA key, but with an attacker-chosen modifier.
  \item \emph{both}: get arbitrary authenticated pointers for a context with attacker-chosen modifier, and the same \PA key.
\end{enumerate}

To prevent the attacker from generating arbitrary authenticated pointers, the program must not contain \PA creation instructions with attacker controlled inputs.
Also, a control-flow attack could be mounted by chaining together instruction sequences to prepare the \PA operand registers with attacker controlled input and then jump to a \PA instruction at another part of the program.
This suggests that \PA-based defenses \emph{must provide, or be combined with, CFI guarantees} that prevent the use of individual authentication instructions as attacker-controlled gadgets.

\paragraph*{Reuse attacks.}

The attacker can read authenticated pointers (including PAC values), and later reuse them to either:

\begin{itemize}
  \item \emph{rollback} an authenticated pointer to a previous value, or
  \item \emph{substitute} an authenticated pointer with another using the same PA modifier.
\end{itemize}

For instance, in GCC's return address signing scheme (Section~\ref{sec:back-qualcomm}), the return address is bound to the location of the stack frame by using the current SP value as the PA modifier.
However, the SP value is not necessarily unique to a specific function invocation.
Consequently, an attacker can reuse the  authenticated return addresses value from one function when a different vulnerable function executes with a matching SP value.
Given that typical programs offer no guarantees on the uniqueness of SP values between different function invocations, this approach exposes a large attack surface for pointer reuse attacks.
Therefore, a concern for any \PA-based defense is partitioning authenticated pointers into distinct classes based on different <\PA key, modifier> pairs.

Attackers can reuse only those pointers they can observe (as opposed all possible values a function pointer can take). Even with full read access to memory (and hence the ability to observe any pointer value that has been generated so far), attackers are still limited to authenticated pointer values the program has already generated.

\section{Adversary Model and Requirements}
\label{sec:adversary-model}

\subsection{Pointer Integrity}
\label{sec:pointer-integrity}

Kuznetsov et al.~\cite{Kuznetsov14} introduced the idea of
\emph{code pointer integrity}: ensuring precise memory safety for all
code pointers in a program.
Since control-flow attacks depend on the manipulation of code pointers, guaranteeing code pointer integrity will render \emph{all} control-flow attacks impossible~\cite{Kuznetsov14}.

The notion of \emph{pointer integrity} is generalizable to both code and data pointers.
In Section~\ref{sec:fully-precise-pointer-integrity}, we provide a more rigorous definition of pointer integrity. Intuitively, pointer integrity aims to prevent unintentional changes to pointers while they remain in program memory so that the value of a pointer at the time it is ``used'' (e.g., dereferenced or loaded from memory) is the same as when it was created or stored on memory.
In particular, integrity-protected pointers reference the intended target objects. 
As explained in Section~\ref{sec:run-time-attacks}, all control-flow attacks, all known DOP attacks and many other data-oriented attacks rely on the manipulation of vulnerable pointers. Consequently, ensuring pointer integrity will prevent these attacks.

\subsection{Attacker Capabilities}
\label{sec:adversary-attacker}

To reason about how effectively PA defends against state-of-the-art attacks we assume attacker capabilities
consistent with prior work on run-time attacks~(Section~\ref{sec:run-time-attacks}).
Our adversary model assumes a powerful attacker with arbitrary memory read and write capabilities restricted only by DEP\@.
The attacker can thus read any program memory and write to non-code segments.
We further assume that the attacker has no control of higher privilege levels, i.e., an attacker targeting a user space process cannot access the kernel or higher privilege levels.
Specifically, we assume that the attacker cannot infer the PA keys, as they are in registers not directly readable from user space (Section~\ref{sec:pa}).
We discuss protection of kernel code using PA in Section~\ref{sec:future-work}.
The attacker's ability to read arbitrary memory precludes the use of randomization-based defenses that cannot withstand information disclosure (e.g., address space layout randomization~\cite{Shacham04} or software shadow-stacks~\cite{Abadi09}).
\PA was specifically designed to remain effective even when the entire memory layout of the victim process is known.

\subsection{Goal and Requirements}
\label{sec:goals-and-requirements}

Our goal is to thwart control-flow and data-oriented attacks by preventing the attacker from forging pointers used by a vulnerable program. We identify the following requirements that our solution should satisfy:

\begin{enumerate}[label=\fbox{R\arabic*},ref=\fbox{R\arabic*}]
\itemsep0em
  \item\label{req:pi}\textit{Pointer Integrity}: Detect/prevent the use of corrupted code and data pointers.

  \item\label{req:attack}\textit{PA-attack resistance}: Resist
    attempts to control PAC generation, and pointer reuse attacks.

  \item\label{req:compat}\textit{Compatibility}: Allow protection of
    existing programs without interfering with their normal operation.

  \item\label{req:perf}\textit{Performance}: Minimize run-time and memory overhead and gracefully scale in relation to the number of protected pointers and dereferences/calls.

\end{enumerate}

\section{Design}
\label{sec:design}

\begin{table*}
  \caption{
    For code and data pointers \SHORTNAME{} uses a static PA modifier based on the pointer's \emph{ElementType} as defined by LLVM\@.
    Return address signing uses a 48-bit \texttt{function-id} and the 16 most-significant bits of the SP value.
  }
  \label{tbl:compartmentalization}
  \centering
    \begin{tabular}{clccl}
      \toprule{}
      & & key & Modifier type & Modifier construction \\
      \midrule{}
      \dCOne{}    & Data pointer signing & Data A & static & \texttt{type-id} = SHA3(ElementType) \\
      \dCTwo{}    & Code pointer signing & Instr A  & static & \texttt{type-id} = SHA3(ElementType) \\
      \dCThree{}  & Return address signing & Instr B & dynamic + static & SP | \texttt{function-id} = compile-time nonce\\
      \bottomrule{}
    \end{tabular}
\end{table*}

To meet our requirements (Section~\ref{sec:goals-and-requirements}) we must solve a number of challenges which we elaborate below\ifabridged.\else:

\it
\begin{enumerate}
  \itemsep0em
  \item[\ref{sec:design-instrument}] \emph{\nameref{sec:design-instrument}}
  \item[\ref{sec:design-create-PACs}] \emph{\nameref{sec:design-create-PACs}}
  \item[\ref{sec:design-pointer-comp}] \emph{\nameref{sec:design-pointer-comp}}
  \item[\ref{sec:design-data-optimization}] \emph{\nameref{sec:design-data-optimization}}
  \item[\ref{sec:design-pointer-conversion}] \emph{\nameref{sec:design-pointer-conversion}}
\end{enumerate}
\normalfont{}
\fi

\subsection{Instrument program with PA instructions}
\label{sec:design-instrument}

To meet requirement~\ref{req:pi}, the program executable must be instrumented with \PA{} instructions to create and authenticate PACs when needed.
For this, we designed and implemented \emph{\LONGNAME} (\SHORTNAME), a compiler enhancement that emits \PA{} instructions to sign pointers in memory as required.
Specifically, it protects:

\begin{itemize}
  \item return addresses;
  \item local, global and static pointers; and
  \item pointers in C structures.
\end{itemize}

\ifnotabridged{}
\begin{figure}[tp]
\centering
  \includegraphics[width=0.8\columnwidth]{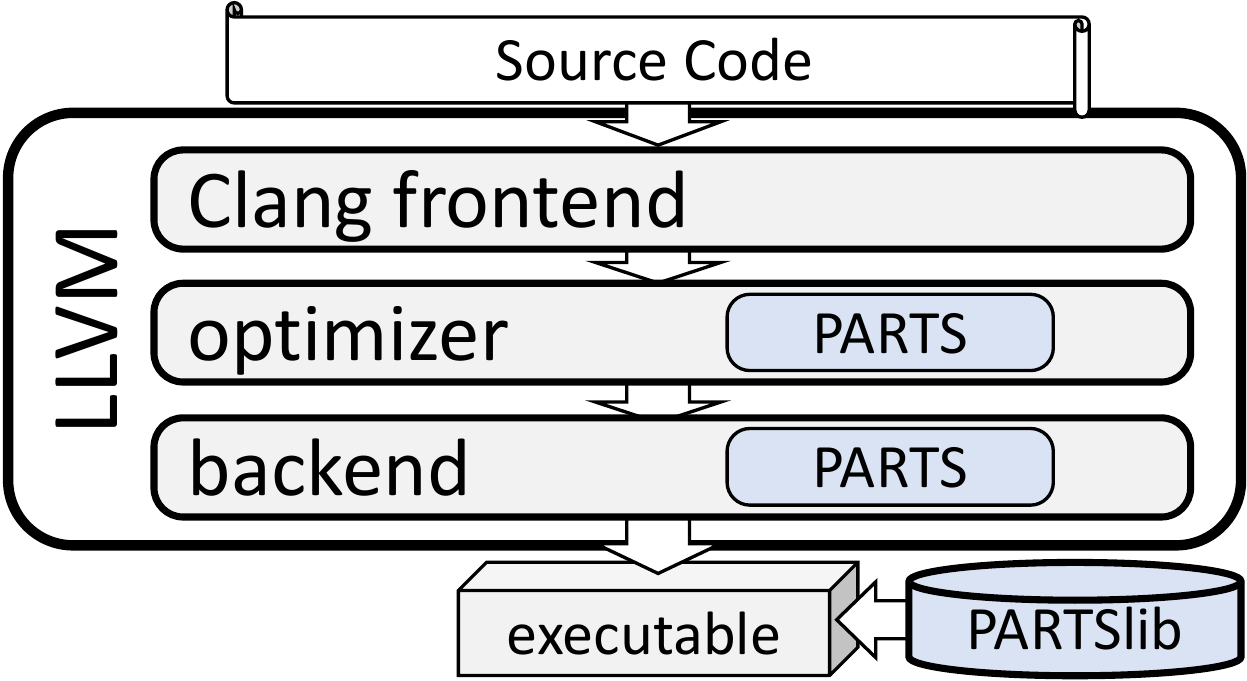}
\caption{
The \SHORTNAME{} instrumentation is setup with compiler modifications and utilizes a run-time support library.
}
\label{fig:parts-overview}
\end{figure}
\fi

Figure\ifabridged~\ref{fig:compilation}\else~\ref{fig:parts-overview}\fi\xspace{} shows the overall architecture of the \SHORTNAME-enhanced compiler.
\SHORTNAME{} analyzes the compiler's intermediate representation (IR) to identify any pointers used by the program and then emits PA instructions at points in the program where pointers are (a) created or stored in memory, and (b) loaded from memory or used.

\subsection{Create PACs in statically allocated data}
\label{sec:design-create-PACs}

Programs may contain pointers which are initialized by the compiler, e.g., defined global variables.
However, PAC values for authenticated pointers cannot be calculated before program execution, as \PA keys are set only at program launch.
Consequently, initialized pointers in the program's data segment pose a challenge, as their values are normally initialized by the linker and loaded into memory separately.
\SHORTNAME{} solves this problem by generating a custom initializer function for pointers requiring PACs.
At run-time, the \SHORTNAME{} runtime library, \SHORTNAMELIB{}, processes the relocated variables and invokes the generated initializer function to ensure that any defined pointers are furnished with a PAC\@.

\subsection{Pointer compartmentalization}
\label{sec:design-pointer-comp}
\label{sec:rts}

As described in Section~\ref{sec:attacks-on-pa} the attacker may attempt to reuse previously signed pointers.
To meet requirement \ref{req:attack} \SHORTNAME{} therefore limits the scope of such reuse attacks by compartmentalizing pointers in three different ways, as shown in Table~\ref{tbl:compartmentalization}.

\noindent\textit{
Code / Data Pointer Compartmentalization:} Recall from Section~\ref{sec:pa}, that \PA provides separate key sets for data and code pointers making it possible to limit reuse attacks.

\noindent\textit{
Run-time type safety:}
Pointer compartmentalization, while effective, is coarse-grained.
To address this, \SHORTNAME{} adds run-time type safety for data and code pointers.
Run-time type safety records the pointer's type by encoding it in the PA modifier.
Then, it checks that pointer dereferences or indirect calls take place using a pointer with a recorded type that matches the type expected at the use site.
\SHORTNAME{} assigns pointers a unique id, \texttt{type-id}, based on the pointer's LLVM \emph{ElementType} which depends on the pointed-to data, structure, or function signature.
Two pointers are \emph{compatible} (have the same \texttt{type-id}) if their ElementType is the same.
\SHORTNAME{} uses a deterministic scheme, detailed in Section~\ref{sec:compiler-type-id} and shown in Table~\ref{tbl:compartmentalization}, to calculate \texttt{type-id}s during compilation.
This ensures that separate compilation units generate equivalent \texttt{type-id}s for compatible objects, and different \texttt{type-id}s for non-compatible ones.

\noindent\textit{
Improved Return Address Signing:}
While run-time type safety could also be applied for return addresses, it would result in an over-permissive policy for backward edges.
As described in Section~\ref{sec:attacks-on-pa}, binding the authenticated return address to the current stack pointer value alone is insufficient because the stack pointer may not be unique to a specific function invocation.
Instead, \SHORTNAME{} uses a combination of the current stack pointer value, and a compile-time nonce (\texttt{function-id}) ensuring that the authenticated return address cannot be reused across invocations of \emph{different functions}, while the stack pointer values effectively compartmentalizes return addresses to callers with different stack layouts.

\subsection{On-load data pointer authentication}
\label{sec:design-data-optimization}

Pointers with PACs can be authenticated either as they are loaded from memory, or immediately before they are used. We refer to these as \emph{on-load} and \emph{on-use} authentication, respectively.
Data pointers are often dereferenced frequently without intervening function calls, i.e., they will not be cleared after use.
This allows the compiler to optimize memory accesses such that, for instance, temporary values might never be written to memory.
\SHORTNAME{} accommodates this behavior by only using on-load authentication for data pointers.
The combined \PA instructions can be used for on-use authentication of code pointers, which are typically loaded to a register, used once, and cleared.
On-load authentication always uses the standalone authentication instructions.
An attacker could attempt to exploit either the standalone authentication or the separate pointer dereference by diverting control flow to either.
However, as mentioned in Section~\ref{sec:attacks-on-pa}, PA solutions must be combined with CFI guarantees, which prevent this type of attacks.

\subsection{Handling pointer conversions}
\label{sec:design-pointer-conversion}

A data pointer to an object of a specific type may be converted to a pointer to a different object type.
When run-time type safety is applied to authenticated pointers, special care must be taken to not interfere with legitimate pointer conversions to meet requirement \ref{req:compat}.
For instance, if a struct pointer is cast to a pointer to its first field, it will change the \texttt{type-id} and hence the expected PAC\@.

If the source and destination object types are compatible, no special consideration is needed.
If not, \SHORTNAME{} must convert the authenticated pointer to the correct \texttt{type-id}.
Because data pointer PAC creation and authentication is done at store/load, \SHORTNAME{} handles conversions by;
\begin{inparaenum}[(a)]
\item if loading the pointer from memory, validating and stripping the PAC using the \texttt{type-id} of the original object, and
\item on store, creating a new PAC using the destination object \texttt{type-id}.
\end{inparaenum}

A pointer to a function of one type may be converted to a pointer to a function of another type.
However, the behavior when calling a function pointer cast to a non-compatible type is undefined~\cite{C1X}[6.3.2.3§8].
Hence, \SHORTNAME{} does not need to convert the pointer's PAC to match the destination function's \texttt{type-id}.
If the converted pointer is converted back, the result is expected to be the same as the original pointer~\cite{C1X}[6.3.2.3§8].
\SHORTNAME{} satisfies this as it does not modify the pointer's PAC\@.

\section{Implementation}
\label{sec:implementation}

\begin{figure}[t]
\centering
\includegraphics[width=0.9\columnwidth]{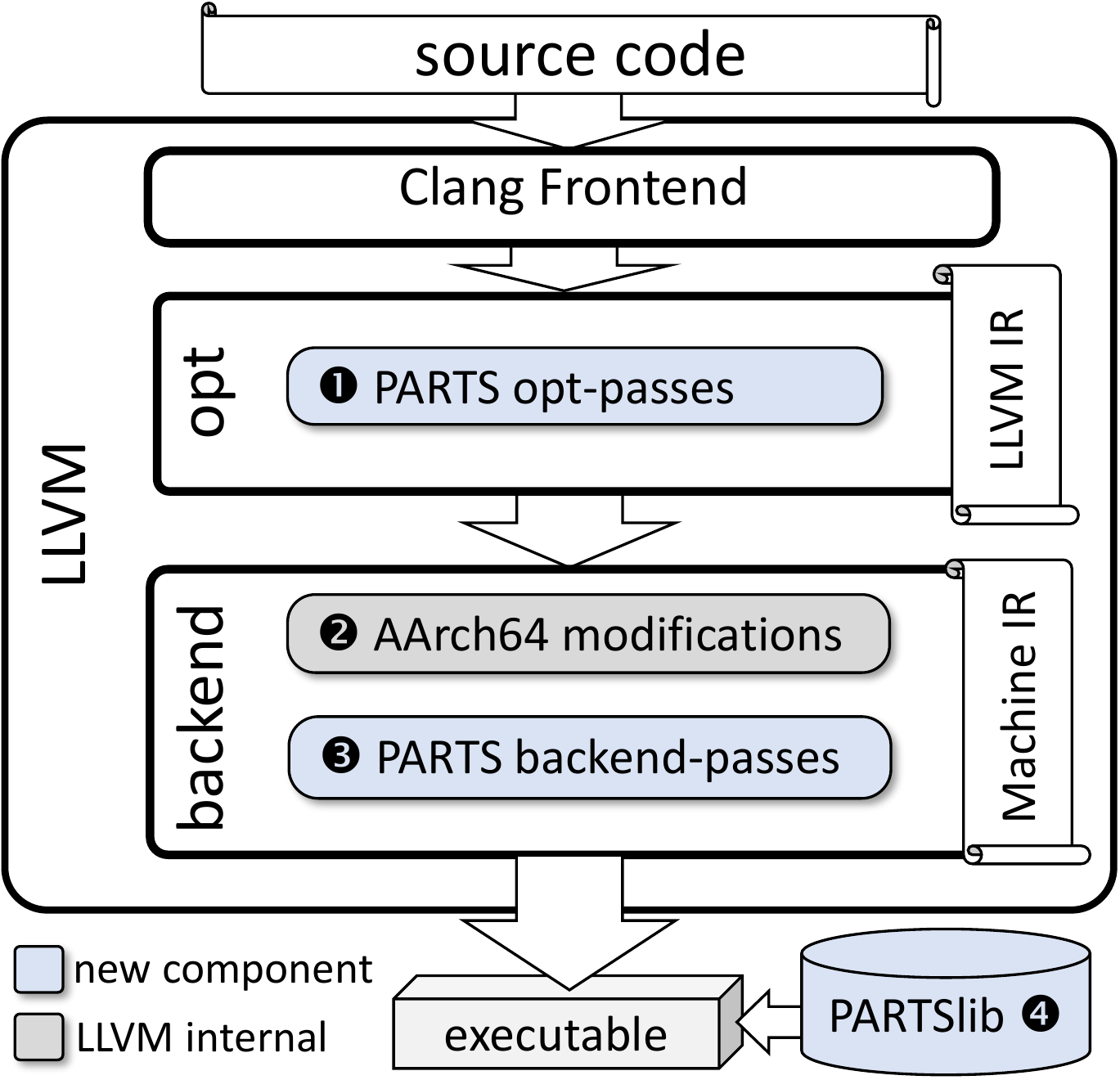}
\caption{
\SHORTNAME{} architecture.
}
\label{fig:compilation}
\end{figure}

The \SHORTNAME{} compiler is based on LLVM 6.0 but modifies and adds new passes to the optimizer and the AArch64 backend (Figure~\ref{fig:compilation}).
The optimization passes~(\dOpt) generate necessary metadata for PA modifiers, inserts wrappers for compatibility with legacy code, and prepares initializers for statically allocated pointers.
The AArch64 Frame Lowering emits function prologues and epilogues and is modified to include instructions for authenticating the LR value~(\dAArch).
The \SHORTNAME{} backend passes~(\dBackend) retrieve the PA modifiers and instruments appropriate low-level instructions.
The resulting binary is linked with \SHORTNAMELIB{}~(\dLib{}), which at run-time creates PACs for the initialized pointers.

\begin{figure*}
  \noindent\begin{minipage}[t]{\columnwidth}
  \begin{lstlisting}[style=customasm,
label={lst:impl-becfi},
caption={
    The \SHORTNAME{} return address signing binds the PAC to the SP~(\dCOne,\dCFive) and unique function id~(\dCTwo,\dCFour).
    The PA modifier is in register \texttt{Xd} during PAC creation~(\dCThree) and authentication~(\dCSix). The 48-bit \texttt{func-id} is split into three 16-bit parts, each moved individually to \texttt{Xd} by left-shifting.
}]
MACRO movFunctionId Mod
  movk  Mod, #func_id16, lsl #16
  movk  Mod, #func_id32, lsl #32
  movk  Mod, #func_id48, lsl #48
ENDM

function:
  mov       Xd, SP            ; (*@\dCOne@*)get SP
  movFunctionId Xd            ; (*@\protect{\dCTwo@*)get id
  pacib     LR, Xd            ; (*@\dCThree@*) PAC
  stp       FP, LR, [SP, #(*@\texttt{0}@*)]  ; store
  ; function body
  ldp       FP, LR, [SP, #(*@\texttt{0}@*)]  ; load LR
  mov       Xd, SP            ; (*@\dCFive@*)get SP
  movFunctionId Xd            ; (*@\protect{\dCFour@*)get id
  autib     LR, X             ; (*@\dCSix@*)auth
  ret
\end{lstlisting}
\end{minipage}\hfill
\begin{minipage}[t]{\columnwidth}
\begin{lstlisting}[style=customasm,
label={lst:impl-code-pointers},
caption={
    The \SHORTNAME forward-edge code pointer signing uses the code pointer's \texttt{type-id} as the PA modifier~(\dOne,\dThree). The 64-bit \texttt{type-id} is split into four 16-bit parts. The PAC is created only once when initially creating the code pointer~(\dTwo). Upon use, i.e., indirect call, the PAC is authenticated using the combined branch and authenticated instruction~(\dFour). \SHORTNAME{} does not instrument intermediate store/load operations.
}]
MACRO movTypeId Mod
  mov   Mod, #type_id00
  movk  Mod, #type_id16, lsl #16
  movk  Mod, #type_id32, lsl #32
  movk  Mod, #type_id48, lsl #48
ENDM

mov       cPtr, #instr_addr  ; load cPtr
movTypeId Xd                 ; (*@\dOne@*)get id
pacia     cPtr, Xd           ; (*@\dTwo@*)PAC
; no intermediate cPtr instrumentation
movTypeId Xd                 ; (*@\dThree@*)get id
blraa     cPtr, Xd           ; (*@\dFour@*)branch
\end{lstlisting}
\end{minipage}
\end{figure*}

\subsection{LLVM Compiler Integration}
\label{sec:compiler-integration}

While the LLVM 6.0 AArch64 backend recognizes PA instructions, they are not used by any pre-existing security feature.
Our modifications consist of added optimizer and backend passes, minor modifications to the AArch64 backend, and new \SHORTNAME{}-specific intrinsics.
Where applicable, we use optimizer passes that operate on the high-level LLVM \emph{intermediate representation} (IR).
Nonetheless, much of the needed functionality is PA-specific and thus implemented in the backend that uses low-level LLVM \emph{machine IR} (MIR), and a register- and instruction set specific to 64-bit ARM\@.

\paragraph*{Determining pointer \texttt{type-id}.}
\label{sec:compiler-type-id}

The compiler backend views the program from a low-level perspective, and the MIR has lost much of the semantics present in C or the high-level IR\@.
Therefore, \SHORTNAME{} must determine \texttt{type-ids} during its optimizer passes where this information is still available (Figure~\ref{fig:compilation},~\dOpt).
The \texttt{type-id} for data consists of a truncated 64-bit SHA-3 hash of the pointer's LLVM \texttt{ElementType}.
The \texttt{ElementType} represents the IR level data type and distinguishes between basic data types, but does not retain \texttt{typedef} or other information from the frontend (i.e., clang).
Code pointers use the same scheme wherein the \texttt{ElementType} consists of the function signature at the same abstraction level.
The \texttt{type-ids} are passed to the backend either via \SHORTNAME{}-specific compiler intrinsics, or by embedding them as metadata in the existing IR instructions.
The AArch64 instruction selection retrieves the information from the IR instructions and transfers it to the emitted MIR~(Figure~\ref{fig:compilation},~\dAArch).
To facilitate the run-time bootstrap (Section~\ref{sec:bootstrap}) \SHORTNAME{} also includes a pass that prepares a custom initializer function that is called at run-time to generate PACs for defined global pointers~(Figure~\ref{fig:compilation},~\dOpt).

\paragraph*{Return addresses signing.}

Return address signing is implemented in the AArch64 backend during frame lowering (Figure~\ref{fig:compilation},~\dAArch).
Frame lowering emits the function prologues and epilogues, and for non-leaf functions, emits instructions for storing and retrieving the LR value from the stack.
\SHORTNAME{} authenticates the value of the LR only if it was retrieved from the stack.
The PAC modifier is based on the 16 least-significant bits of the SP value and a 48-bit function-specific \texttt{function-id}.
The \texttt{function-id} is guaranteed to be unique within the current compilation unit or, with link time optimization (LTO), the whole program.
To avoid repetition across different compilation units, the \texttt{function-id} is generated using a pseudorandom, non-repetitive sequence.

\paragraph*{Code pointer signing.}

\SHORTNAME{} uses the combined PA instructions for branches and converts branch instructions directly to their PA variants (Figure~\ref{fig:compilation},~\dBackend).
The PAC for any code pointer is created only once at the time of pointer creation, e.g., when the address of a function is taken.
This is instrumented by adding a PAC-creation instruction immediately after the instruction that moves a code pointer to a register.
Subsequent load and store operations do not authenticate the signed code pointers, instead they are authenticated only on use.

\paragraph*{Data pointer signing.}

As discussed in Section~\ref{sec:design-data-optimization}, it is not feasible to perform on-use authentication for data pointers.
Instead, we authenticate data pointers when they are loaded from memory and create PACs before storing them.
In some cases, e.g., using globals, the IR will include explicit load and store operations that can be furnished with the \texttt{type-id}.
Our modified Instruction Selection then forwards the \texttt{type-id} to the emitted MIR~(Figure~\ref{fig:compilation},~\dAArch).
However, stack-based store and load operations, in particular, are often not present before the backend finalizes the stack-layout and register allocation.
Thus, some load and store instructions must be instrumented solely in the backend.

While it would be possible to modify the AArch64 backend (e.g., register allocation), we have instead opted for a less invasive approach.
The \SHORTNAME{} backend pass~(Figure~\ref{fig:compilation},~\dBackend{}) finds load and store instructions in the MIR, and uses the attached \texttt{type-id} for instrumentation.
When the \texttt{type-id} is not present, e.g., because the load and store is a register spill, the \texttt{type-id} is fetched from surrounding code.
For instance, when instrumenting the store due to register spilling a pointer variable, the correct \texttt{type-id} can be fetched from the original load.

\subsection{Run-time Bootstrap}
\label{sec:bootstrap}

Programs may contain pointers in statically allocated data, i.e., pointers stored in global variables or static local variables.
These are initialized by the compiler or linker, and therefore cannot include PACs.
The \SHORTNAMELIB{} runtime library instead invokes the compiler generated custom PAC initializer function at process startup.
Our Proof-of-Concept implementation invokes the \SHORTNAMELIB{} bootstrap using compiler instrumentation that explicitly calls the functionality when entering \texttt{main}.

\ifnotabridged{}
Our current approach relies on LTO, because the initializer function is created once for each optimization unit.
An alternative is to use the C constructor feature supported by Clang and GCC\@.
The libc initialization and will run all constructor functions before invoking the program's main function.
The order in which constructor functions are run is well-defined only within the same translation unit.
This means that programs that already use C constructors to run custom code may interfere with the PA initialization routine.
Therefore, we aim to move support for PA directly to the dynamic linker (see Section~\ref{sec:future-work}).
\fi

\subsection{Instrumentation}
\label{sec:instrumentation}

\SHORTNAME{} uses only in-line instrumentation and does not require storage of separate run-time metadata.
With the exception of the bootstrap process the original code structure is thus largely unchanged.
As discussed in Section~\ref{sec:pa}, no explicit error handling is added by \SHORTNAME{};
instead, an authentication failure will set specific high-order bits in the pointer, thus triggering a memory translation fault on subsequent dereference or call using the pointer that failed authentication.
The high-order bits ensure that the fault is distinguishable as one caused by authentication failure.
Our code listings use two macros for setting up PA modifiers for return address signing and \texttt{type-id} based PACs, these are shown in Listing~\ref{lst:impl-becfi} and Listing~\ref{lst:impl-code-pointers}.

\paragraph*{Return address signing.}

The return address signing instrumentation is similar to GCC's implementation~\cite{Qualcomm17} but includes an added modifier (Listing~\ref{lst:impl-becfi}).
The function prologue is instrumented such that it prepares the PA modifier by moving SP~(\dCOne) value into a free register.
The SP value is combined with the \texttt{function-id}~(\dCTwo) to form the PA modifier, which is then used with the instruction B key~(\dCThree).
The \texttt{function-id} is generated at compile-time using LLVM's random number generator, and is guaranteed to be unique withing the LLVM Module (i.e., the whole program, when using link time optimization).
The function epilogues (i.e., any part that ends with a return or a tail-call) are similarly instrumented to generate the same PA modifier~(\dCFour,\dCFive) and to verify the PAC in the restored LR~(\dCSix).

\paragraph*{Code pointer signing.}

\SHORTNAME{} instruments code pointers only on creation and use (Listing~\ref{lst:impl-code-pointers}).
Specifically, when a code pointer is initially created, \SHORTNAME{} will use the instruction A-key to create a PAC~(\dTwo) based on the target \texttt{type-id}~(\dOne).
The instrumentation will at no point remove the PAC from a code pointer.
Instead, \SHORTNAME{} uses the combined authenticate and branch instructions --- e.g., \texttt{blraa} --- to perform the branch directly on an authenticated pointer~(\dFour), again using the same PA modifier~(\dThree).

\paragraph*{Data pointer signing.}

All data pointer stores and loads are instrumented such that a PAC is created immediate before store and authenticated immediately after load (Listing~\ref{lst:impl-data-pointers}).
When a data-pointer is used the instrumentation first sets up the correct PA modifier, i.e., the \texttt{type-id}~(\dCOne).
The pointer is then immediately authenticated using the modifier and data A-key~(\dCTwo); this also strips the PAC from the pointer.
As long as the data pointer resides in a register it can thus be used without any performance overhead.
\SHORTNAME{} creates PACs for pointers immediately before store in the same manner, save for the \texttt{pacda} instruction.

\begin{lstlisting}[float,floatplacement=tp,style=customasm,
label={lst:impl-data-pointers},
caption={
    \SHORTNAME{} immediately authenticates data pointers loaded from writeable memory.
    This is done by first loading the \texttt{type-id}~(\dCOne) and then verifying the PAC~(\dCTwo).
}]
ldr   dPtr, [SP, #(*@\texttt{0}@*)]   ; load dPtr
movTypeId Xd, #type_id ; (*@\dCOne@*)get id
autda dPtr, Xd         ; (*@\dCTwo@*) authenticate
; dPtr is directly usable
\end{lstlisting}

\section{Evaluation}
\label{sec:evaluation}

We develop our Proof-of-Concept implementation of \SHORTNAME{} on the ARMv8-A Base Platform Fixed Virtual Platform (FVP), based on Fast Models 11.4, which supports version 8.0 to 8.4 of the ARMv8-A architecture~\cite{ARMFVP}.
At the time of writing, the only PA-capable hardware is the Apple A12 and S4 SoCs featuring ARMv8.3-A CPUs~\cite{iOSSecurity}.
However, these proprietary SoCs are, to the best of our knowledge, not available in development versions outside Apple.
The FVP provides a software simulation of an ARMv8.3-A processor in AArch64 mode, and is, to the best of our knowledge, the only publicly available environment with \ARMPAFULL{} support\@.

\subsection{ARMv8.3 Emulation and Software Stack}
\label{sec:emulation}

\begin{figure}[th]
\centering
\includegraphics[width=0.8\columnwidth]{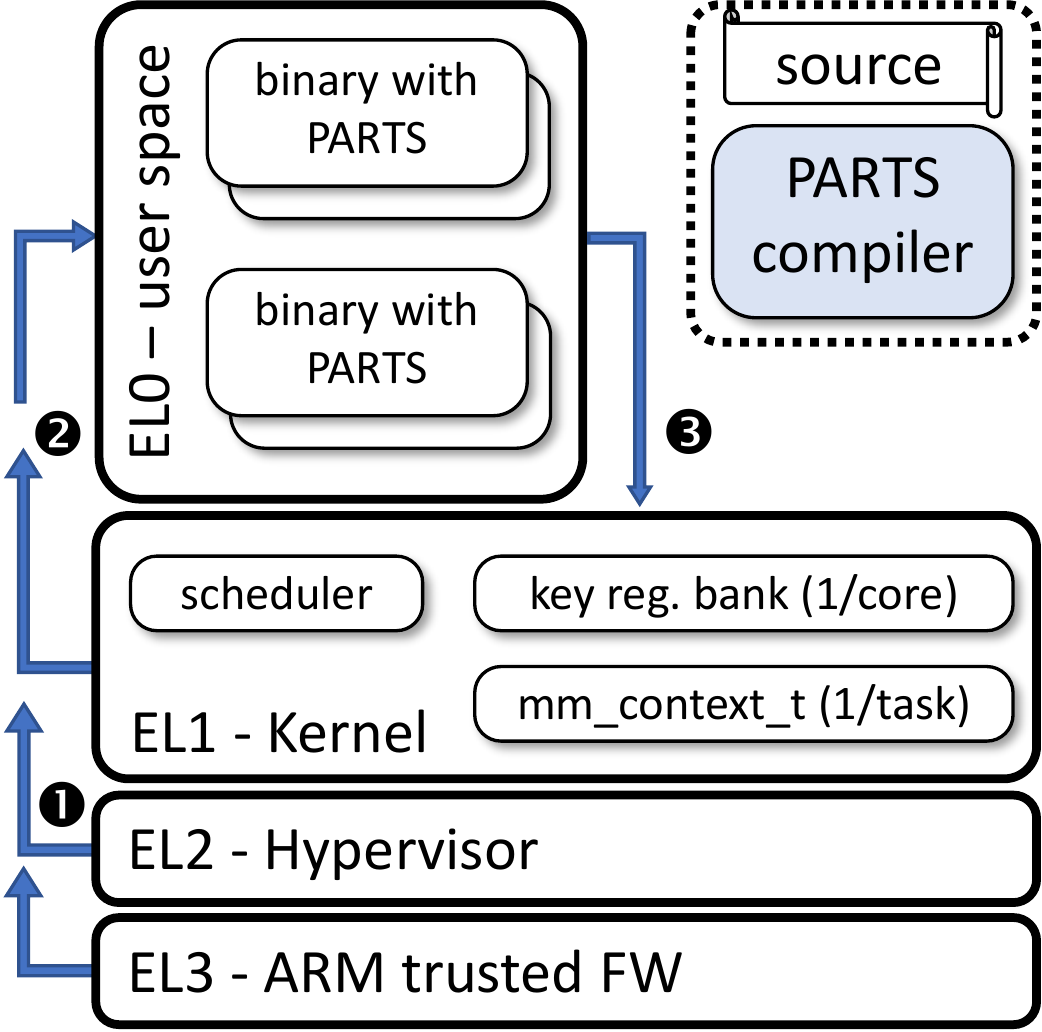}
\caption{The trapping of PA configuration must be released \dOne, in order to allow the kernel to manage the PA keys on process creation and context switches \dTwo. Faults generated by failed authentications will be trapped by the kernel \dThree.}
\label{fig:architecture}
\end{figure}

We use GNU/Linux with a 4.14 kernel, modified to support \PA\@.
We modified the bootloader and kernel to activate \ARMPAFULL, and allow key configuration during kernel scheduling at Exception Level 1 (EL1 in Figure~\ref{fig:architecture}).
Our kernel modifications are based on Mark Rutland's 2018 \PA{} patches\footnote{\url{https://lwn.net/Articles/752116/}}.

\ifnotabridged{}

During system boot, the PA-setup proceeds as follows:
As the PA feature is turned off by default it needs to be activated, through the \texttt{SCR\_EL3} system control register, by the ARM Trusted Firmware, i.e., at Exception Level 3 (EL3).
Second, in order to support kernel (EL1) or hypervisor (EL2) use, the hardware by default traps the use of PA and the setting of PA keys to EL3.
As a consequence, we must release the trapping of PA instructions to these exception levels (bits \texttt{SCR\_EL3\_API}, \texttt{HCR\_EL2\_API} in the \texttt{SCR}/\texttt{HCR} registers respectively) as well as the trapping of PA key writing (bits \texttt{SCR\_EL3\_APK}, \texttt{HCR\_EL2\_APK}).
The PA key management in the Linux kernel can only be performed with these preconditions.

The kernel scheduler is modified to dynamically determine whether \PA{} is enabled.
\fi
\PA{} keys for each task are stored in a process-specific \texttt{mm\_context\_t} structure (in the process' memory descriptor in the kernel) which contains architecture-specific data related to the process address space.
Threads within the same process have a common memory descriptor, and thus share the same \PA{} keys.
The scheduler will configure the \PA{} key registers using the keys in the process' memory descriptor whenever a task is scheduled to run.
When a new child process is forked, the parent's keys are duplicated to the child's memory descriptor. However, when a new executable file is \texttt{exec}'d in the context of an existing process, the kernel initializes a new set of \PA{} keys using \texttt{get\_random\_bytes()}.
In other words, each new process receives a new set of PA keys which remain unchanged thereafter.

\subsection{Security Evaluation}
\label{sec:sec-eval}

\ifnotabridged
\SHORTNAME{} provides a practical realization of Pointer Integrity.
\SHORTNAME{} ensures:
\begin{inparaenum}[1)]
\item that code pointers in indirect branches are always authenticated using the combined branch and authenticate instructions, and
\item that a data pointer dereference is always preceded by a PA authentication, if the data pointer was read from memory.
\end{inparaenum}
We evaluate the security properties of \SHORTNAME{} and demonstrate its practical efficacy in preventing existing attacks.
\fi

\subsubsection{Return address signing}

Return address signing in both GCC~\cite{Qualcomm17}, and \SHORTNAME{} prevents an attacker from introducing forged return addresses to the program stack\ifnotabridged\xspace(\ref{req:pi})\fi.
\ifnotabridged{}
\SHORTNAME{} further narrows the scope for reuse attacks compared to the return address signing in GCC (\ref{req:attack}).
Recall that return address signing in GCC~\cite{Qualcomm17} determines a function's execution context solely using the SP value.
Therefore, it falls short of compartmentalizing return addresses to individual function invocations in cases where the value of the SP coincides between different function invocations.
The likelihood of SP values coinciding depends on many factors, including the order in which functions are called, and stack frames sizes of functions.
Determining the possible SP value collisions in an arbitrary program's call graph, in other words the susceptibility of the program towards return address reuse attacks, would require an exhaustive search through all the program's potential stack states.

\fi
Compared to GCC, \SHORTNAME{} augments the PA modifier used for return address signing by combining a function-specific identifier with the SP value\ifabridged\xspace(\ref{req:attack})\fi.
As a result, \SHORTNAME{} return address signing precludes the possibility of reuse of the return address between different functions, irrespective of SP value collisions.
It remains susceptible to pointer reuse between distinct invocations of the same function from call sites with same SP value\ifabridged\xspace(\ref{req:pi})\fi.

\subsubsection{Forward-edge code pointer signing}

As with \SHORTNAME{} return address signing, forward-edge code pointer signing prevents an attacker from using forged code pointers injected into program memory (\ref{req:pi}).
This prevents a large class of attacks (e.g., typical ROP/JOP gadgets) that rely on redirecting the control flow to code in the middle of functions, i.e., addresses that never were valid targets of benign control-flow transfers.

\SHORTNAME{} restricts forward-edge code pointer reuse by enforcing \emph{run-time type safety} for signed pointers (\ref{req:attack}).
Under this scheme, pointers used in a pointer reuse attack must share the same \texttt{type-id} (i.e., have a matching type on the LLVM IR level).
This prevents large classes of function-reuse attacks.
The solution is compatible with common programming patterns involving function pointers (\ref{req:compat}), such as callbacks, but allows reuse between code pointers to functions with identical type signatures.

\subsubsection{Data pointer signing}

\SHORTNAME{} data pointer signing protects all data pointers and prevents an attacker from loading a forged data pointer to program memory (\ref{req:pi}).
This prevents all non-control data attacks that rely on corrupting data pointers to unintended parts of of memory. This class of attacks includes all currently known DOP attacks~\cite{Hu16}.

\SHORTNAME{} restricts data pointer reuse by enforcing run-time type safety also for data pointers (\ref{req:attack}).
Reuse attacks would be more useful to an attacker if they could substitute a vulnerable pointer with one referencing an object of different size or type. Therefore restricting pointer substitution based on the pointer's type restricts the attacker's capability to cause unintended data flows within the program.
However, pointer conversions are a challenge for data pointer integrity.
As discussed in Section~\ref{sec:rts}, \SHORTNAME{} accommodates data pointers that are cast from type $A$ to an incompatible type $B$ by writing the converted pointer using the \texttt{type-id} of $B$.
This may expand the effective set of reusable pointers under our threat model;
the attacker can record pointers of type $A$ and reuse them at PAC conversion site $A \rightarrow B$, thereby obtaining a pointer of type $B$ to an object of type $A$.
This converted pointer can then be used at de-reference sites that require pointers of type $B$.
If the program also includes a conversion from $B$ to $A$ this makes both types interchangeable.

\SHORTNAME{} data pointer integrity does not guarantee spatial safety of pointer accesses to data objects, nor does it address the temporal safety (e.g., prevent use-after-free conditions). ARMv8-A PA does not provide facilities to directly address these challenges. We discuss orthogonal schemes that can be used in combination with \SHORTNAME{} to provide spatial and temporal safety guarantees in Section~\ref{sec:related-work}.

\subsubsection{PAC entropy}
\label{sec:eval-pac-entropy}

As explained in Section~\ref{sec:attacks-on-pa}, the PAC size $b$ is a concern for any PA-based scheme.
On typical AArch64 Linux systems, $b$ is between 16 and 24.
To succeed with probability $p$, a PAC guessing attack requires
$\frac{\log(1-p)}{\log(1-2^{-b})}$ guesses on the assumption that a PAC comparison failure leads to program termination.
On our simulator setup where $b=16$, achieving a $50$\%-likelihood for
a correct guess requires $45425$ attempts.

Note that ROP/DOP attacks require an environment where a set of jumps (gadgets) can be set up, each requiring a separate PAC to be broken.
Consequently, success probability of a complete attack will decrease exponentially with the number of jumps necessary.

Pre-forked or multithreaded programs will share the same PA key
between the parent and all sibling threads/processes.
This could allow an attacker to brute force a PAC by targeting a sibling, if PAC failure on a sibling does not result in the termination (and hence PA key reset) of all threads/processes sharing the same PAC key.
In this scenario, $2^{b-1}$ guesses on average are enough to guess a $b$-bit PAC ($32768$ guesses for $b=16$). 
Multithreaded / pre-forking applications could be hardened against guessing attacks by requiring a full application restart if the number of unexpected terminations of child threads/processes exceeds a pre-defined threshold.

\subsection{Performance Evaluation}
\label{sec:perf-eval}

The FVP processor, peripheral models, and micro-architectural fabric is simplified.
Consequently, timing on the FVP model differs from actual hardware.
The ARM Fast Models documentation states that \textit{''all instructions execute in one processor master clock cycle``}.
We confirm this behavior for PA instructions in the FVP by using microbenchmarks that allow PA instructions to be timed in isolation\ifnotabridged\ (Section~\ref{sec:microbenchmarks})\fi.
As a result, we cannot use the FVP to estimate the expected run-time overhead of \SHORTNAME{}.
Instead, we estimate the execution time of PA instructions and develop a PA-analogue that emulates the run-time cost of PA instructions (Section~\ref{sec:eval-pa-analogue}).
We then run large-scale benchmarks on real (non-PA) hardware using our PA-analogue (Section~\ref{sec:eval-nbench}).

\ifnotabridged{}
\subsubsection{Confirming simulator behavior}
\label{sec:microbenchmarks}

To measure PA instruction performance on the FVP we use a hand-crafted assembly loop with a \instr{pacia}~(\dCTwo) and \instr{autia}~(\dCThree) instruction (Listing~\ref{lst:bench-instr}).
We configured this loop to execute $10^7$ times~(\dCOne) and measured run time by reading the \textit{cntvct\_el0} register, which provides access to a timer clocked at 100MHz.
We then compared the timing with and without the PA instructions~(\dCTwo,\dCThree).
To exclude potential differences between different host machines we took measurements with FVP rate limiting enabled on the following underlying host CPUs: i7--8700K, i7--7600U and i7--7500U.
In all cases, we observed an overhead of $50.00\%$, which is consistent with the assumed PA instruction behavior on the FVP\@.

\begin{lstlisting}[float,floatplacement=tp,style=customasm,
label={lst:bench-instr},
caption={
Minimal loop for timing PA instructions on FVP. Results and documentation indicate that all instructions require only one cycle, i.e., in this case adding two PA instructions among the four non-PA instructions incurs a $50\%$ overhead.
}]
00000000000008b0 <start>:
 8b0: eb0a011f  cmp   x8, x10      ; (*@\dCOne@*)
 8b4: 540000aa  b.ge  8c8 <exit>
 8b8: dac10109  pacia x9, x8       ; (*@\dCTwo@*)
 8bc: dac11109  autia x9, x8       ; (*@\dCThree@*)
 8c0: 91000508  add   x8, x8, #0x1
 8c4: 17fffffb  b     8b0 <start>
\end{lstlisting}
\fi

\subsubsection{PA-analogue}
\label{sec:eval-pa-analogue}

From~\cite[Table 8]{Avanzi17} we can deduce that on a (1.2GHz) mobile core, the PAC is computable with an approximate overhead of 4 cycles, without accounting for the potential speed benefits of opportunistic pipelining or the inclusion of several parallel PAC computing engines per core. For simplicity, we assume equal cycle counts for all PA instructions.
Based on this assumption we construct a \emph{PA-analogue} (Listing~\ref{lst:pa-analogue}) as a proxy to measure overhead of PA instrumentation on non-PA CPUs: it consists of four exclusive-or (\texttt{eor}) operations to account for the 4 cycles. The final \texttt{eor} operates on the modifier and SP to enforce a memory read/write dependency, thus preventing the CPU pipeline from arbitrarily delaying the operations.
We have confirmed that our PA-analogue exhibits the expected overhead using our microbenchmarks.

\begin{lstlisting}[float,floatplacement=tp,style=customasm,
label={lst:pa-analogue},
caption={
PA-analogue simulating PA instructions
}]
 eor  Xptr, Xptr, #0x2 ; spend cycles
 eor  Xptr, Xptr, #0x3 ; to approximate
 eor  Xptr, Xptr, #0x5 ; PA instruction
 eor  Xptr, Xptr, Xmod ; overhead
\end{lstlisting}

\ifnotabridged{}
\subsubsection{Expected overhead}
\label{sec:eval-micro-level}

Based on expected PA instruction cost we can estimate micro-level overhead (\ref{req:perf}) of \SHORTNAME{}.
In particular, we can estimate which factors will contribute to the overhead for specific \SHORTNAME{} features.
For a single PA instruction our instrumentation overhead consists of four move instruction to prepare the PA modifier, and a PA instruction.
In modern ARM Cortex-A processors certain \instr{movk}/\instr{movk} pairs --- e.g., those used \SHORTNAME{} to prepare the PA modifiers (Listings~\ref{lst:impl-becfi},~\ref{lst:impl-code-pointers}, and~\ref{lst:impl-data-pointers}) --- can be executed with one-cycle execute latency and four-instruction/cycle execution throughput~\cite{ARMoptimization}.
Therefore, we estimate the cost of a single PAC creation or authentication to be between \numrange{6}{8} cycles.
Based on microbenchmarks similar to Listing~\ref{lst:bench-instr}, we have confirmed that our instrumentation, using our PA-analogue, causes micro-level overheads inline with these estimates.
However, the proportional overhead depends on the specific program and enabled \SHORTNAME{} features.

\paragraph{Return address signing.}
Each non-leaf function call requires two PA instructions for storing and loading the return address, for an estimated run-time cost of \numrange{12}{16} cycles.
The total overhead introduced by \SHORTNAME{} return address signing, therefore, scales linearly with the number of non-leaf function invocations.

\paragraph{Forward-edge code pointer integrity.}
Code pointers are instrumented only on initial pointer creation, and on all subsequent indirect calls using the pointer.
For a typical program, the expected overhead will largely consist of authenticated indirect calls; each with an estimated \numrange{6}{8} cycle cost.
Similarly, any new code pointers created at turn-time incur an estimated \numrange{6}{8} cycle overhead when created.

\paragraph{Data pointer integrity.}
All data pointer loads and stores, to/from \emph{any} memory, are instrumented; each with an estimated overhead of \numrange{6}{8} cycles.
In particular, this includes memory stores and loads where intermediary values are temporarily written to the stack (e.g., during function calls).
Compiler optimizations generally strive to minimize intermediate stores into memory, and as a result also reduce the number of \SHORTNAME{} on-load authentications.
In other words, the performance overhead incurred by \SHORTNAME{} data pointer integrity scales linearly with the number of stores and loads, rather than the number of pointer dereferences.
We estimate the worst case impact of \SHORTNAME{} data pointer integrity on synthetic performance benchmarks in Section~\ref{sec:eval-nbench} by disabling all compiler optimizations.
\fi

\subsubsection{nbench-byte benchmarks}
\label{sec:eval-nbench}

For our performance evaluation we use the Linux nbench-byte 2.2.3 synthetic benchmark\footnote{\url{http://www.math.utah.edu/~mayer/linux/bmark.html}} designed to measure CPU and memory subsystem performance, providing a reasonable prediction of real-world system performance\footnote{\url{http://www.math.utah.edu/~mayer/linux/byte/bdoc.pdf}}.
We follow work such as~\cite{Brasser17,Chen17,Lee17,Seo17, Shih17,Chen17} and use nbench rather than the SPEC CPU standardized applications benchmarks for our evaluation, as nbench allows us verify the functionality of \SHORTNAME instrumentation with manageable simulation times on the FVP. The current version of the SPEC CPU benchmark suite, SPEC CPU2017\footnote{\url{https://www.spec.org/cpu2017/}}, has replaced many tests in the previous, now retired SPEC CPU2006\footnote{\url{https://www.spec.org/cpu2006/}} with significantly larger and more complex workloads (up to \textasciitilde10X higher dynamic instruction counts). As a result, the SPEC simulation times on the FVP proved to be unmanageable; for example, running individual SPEC benchmarks take hours to \emph{days} to complete on the FVP. This is a challenge for both researchers and industry practitioners who rely on hardware simulation for evaluation~\cite{Panda18}. We report our results for a subset of SPEC CPU2017 tests in Appendix~\ref{sec:appendix-spec}.

\begin{figure*}
  \centering
  \begin{subfigure}[b]{\textwidth}
    \includegraphics[width=\textwidth]{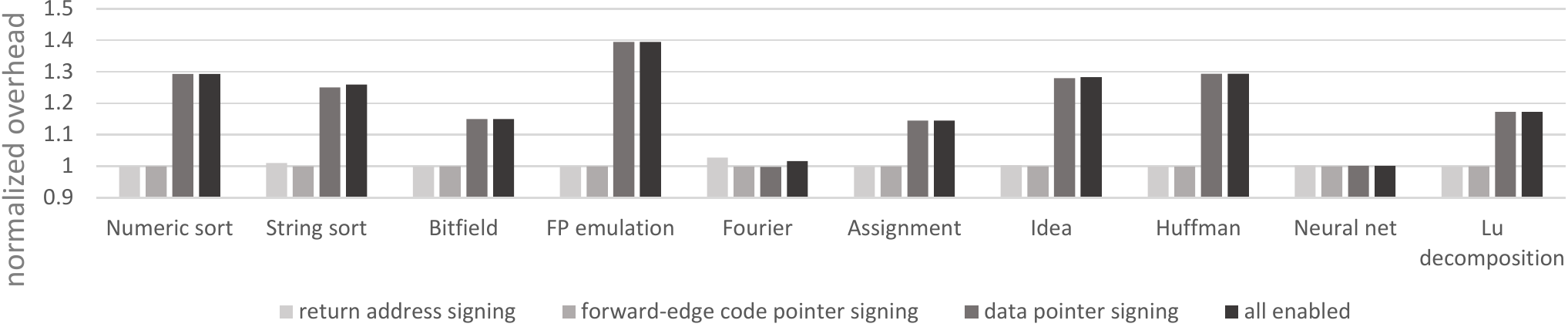}
    \caption{Results of instrumented nbench-byte tests features, normalized to a non-instrumented baseline.}
    \label{fig:nbench_results_a}
  \end{subfigure}
  \begin{subfigure}[b]{\textwidth}
    \includegraphics[width=\textwidth]{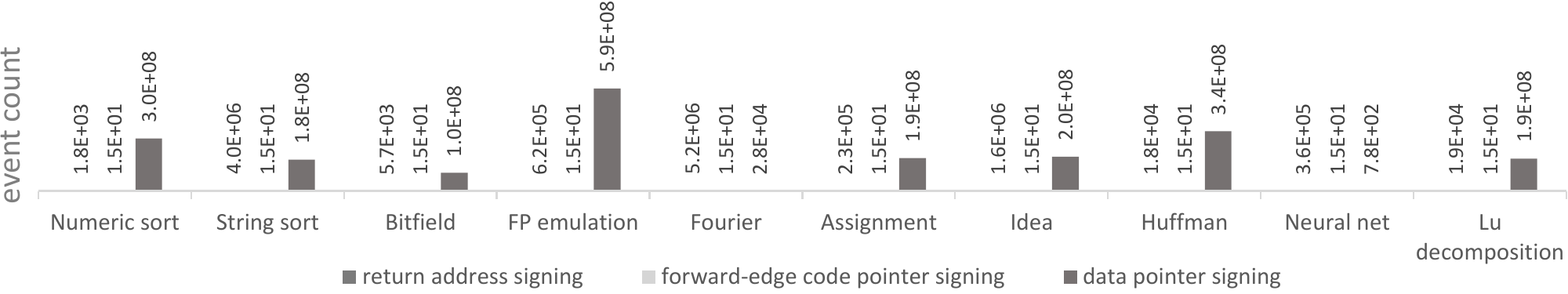}
    \caption{Run-time count of executed locations instrumentable by \SHORTNAME{}.
    Because the program's memory profile affects performance the benchmark results clearly correlate with observed memory use (e.g., FP emulation has a large data pointer integrity overhead because it uses many data pointers)}
    \label{fig:nbench_results_b}
  \end{subfigure}
  \caption{nbench benchmark results}
  \label{fig:nbench_results}
\end{figure*}

The nbench benchmarks include 10 different tests. We adopt the same methodology as Brasser et al.~\cite{Brasser17} and run each test a constant number of iterations for the following cases: 
\begin{inparaenum}[a)]
\item uninstrumented baseline
\item each \SHORTNAME scheme (return address signing, forward-edge code pointer integrity, and data pointer integrity) enabled individually, and 
\item all schemes enabled simultaneously.
\end{inparaenum}
\ifnotabridged{}
All benchmarks are compiled with the LLVM 6.0, but using different switches to enable measured \SHORTNAME{} features.
\fi
Compiler optimizations were disabled for all tests.
The tests were performed on a 96boards Kirin 620 HiKey (LeMaker version) with a ARMv8-A Cortex A53 Octa-core CPU (1.2GHz) / 2GB LPDDR3 SDRAM (800MHz) / 8GB eMMC, running the Linux kernel v4.18.0 and BusyBox v1.29.2.
Figure~\ref{fig:nbench_results} shows the results, normalized to the baseline. A more detailed description can be found in Appendix~\ref{sec:appendix-nbench}.

Return address signing incurs a negligible overhead of less than $0.5$\%.
This is expected because the estimated per-function overhead of \numrange{12}{16} cycles is typically small compared to the full execution time of the instrumented function. The same holds for indirect calls (6-8 cycle overhead at the call site), although indirect calls are underrepresented in nbench-byte.
\ifabridged{}
However, our microbenchmarks for the code pointer integrity instrumentation indicate that a \numrange{6}{8} cycle overhead per indirect function call is reasonable under the assumed QARMA performance.
\else
However, our microbenchmarks for the code pointer integrity instrumentation indicate that the estimate of a \numrange{6}{8} cycle overhead per indirect function call (Section~\ref{sec:eval-micro-level}) is reasonable under the assumed QARMA performance.
\fi

Data pointer integrity\ifnotabridged, as mentioned (Section~\ref{sec:eval-micro-level}), \fi\ depends largely on the memory profile of the instrumented program.
For instance, the floating point emulation test extensively handles data pointers, resulting in a $39.5$\% overhead.
In contrast, the Fourier and neural network benchmarks contain no data pointers and thus incur no discernible overhead.
The geometric mean of the overhead of the combined instrumentation for all tests is $19.5$\%.

\subsection{Compatibility Evaluation}
\label{sec:compat-eval}

Based on our evaluation, \SHORTNAME{} is compatible with standard C code (\ref{req:compat}).
\ifnotabridged{}
However the presence of PACs in protected pointers may interfere with code that expects a particular pointer layout.
This is a limitation of any PA-based solution.
\fi
Because return address signing only affects the instrumented function, it can be safely applied without interfering with the operation of other parts of programs, or uninstrumented code.

\SHORTNAME{} forward-edge code pointer integrity and data pointer integrity can be safely applied to complete code bases.
However, if \SHORTNAME{} is applied only to a partial code base, the instrumented code interfacing with non-instrumented (legacy) libraries requires special consideration.
In particular pointers used by both instrumented and uninstrumented code cannot be passed directly between them. We discuss solutions for backwards compatibility with legacy libraries in Section~\ref{sec:future-work}.

We encountered no compatibility issues with \SHORTNAME{} during our performance evaluation with nbench (Section~\ref{sec:perf-eval}).

\section{Related Work}
\label{sec:related-work}

\emph{Code-pointer integrity (CPI)}~\cite{Kuznetsov14} protects access to code pointers --- and data pointers that may point to code pointers --- by storing them in a disjoint area of memory; the \emph{SafeStack}\footnote{\url{https://clang.llvm.org/docs/SafeStack.html}}.
The SafeStack itself must be protected from unauthorized access.
Randomizing the location of the SafeStack is efficient~\cite{Kuznetsov15}, but easily defeated by an attacker who can read arbitrary memory.
Stronger protection of the SafeStack using hardware-enforced isolation or software-isolation incurs an average performance overhead of 8.4\% or 13.8\% in SPEC CPU2006 benchmarks.
\ifnotabridged{}
\emph{Code-Pointer Separation} (CPS)~\cite{Kuznetsov14} is a variant of CPI that only secures code pointers to achieve reduced run-time overhead.
CPS implemented using hardware-enforced segmentation or information hiding incurs a performance overhead in the order of \numrange{1.8}{2.2}{\%}.
\fi

\paragraph*{Protecting pointers using cryptography.}

Prior cryptographic defenses against run-time attacks generally assume the attacker cannot read memory. \emph{PointGuard}~\cite{Cowan03} instruments a program to apply a secret XOR mask to all pointer values.
This prevents an attacker from reliably forging pointer values without knowledge of the mask.
\emph{Data randomization}~\cite{Cadar08} extends data masking to cover all data in memory.
It uses static points-to analysis and distinct masks to partition memory accesses in separate classes.
\ifnotabridged{}
Both PointGuard and data randomization rely on the secrecy of the XOR mask, but store their secrets within the process' address space.
\fi
Neither PointGuard nor data randomization remain effective under our threat model.

Similarly to \ARMPAFULL, \emph{Cryptographic CFI} (CCFI)~\cite{Mashtizadeh15} uses MACs to protect control-flow data, such as return addresses, function pointers, and vtable pointers. Like \SHORTNAME, CCFI uses a function's type signature to separate function pointers to distinct protection domains, but does not protect function pointers embedded in C structures.
\ifnotabridged{}
The use of MACs gives CCFI and \ARMPA{} several useful advantages over traditional CFI approaches: it prevents attackers from introducing non-authenticated pointers to the program memory, it allows separating pointers into different protection domains based on static, or run-time characteristics, which enables more finer-grained separation of sensitive pointers than stateless CFI\@.
\fi
Unlike \ARMPA, CCFI only benefits from hardware-accelerated AES for speeding up MAC, resulting in a high performance overhead (52\% overhead on average in SPEC CPU2006 benchmarks). In contrast, \SHORTNAME{} also benefits from hardware-accelerated checks by using ARMv8-A PA instructions, protects both code and data pointers, including pointers embedded in C structures.

\paragraph*{Hardware-assisted mechanisms.}

Various hardware-assisted defenses are described in research literature~\cite{Devietti08,Woodruff14,Watson15,Davi15,Song16,Tsampas17,Nyman17b,Roessler18}.
\ifabridged{}
The majority of such defenses have only been realized as soft microprocessor prototypes on FPGAs.
Here we describe mechanisms available in commercial off-the-shelf processor architectures.
\else
\emph{CHERI}~\cite{Woodruff14} is a hardware-assisted memory capability model for the 64-bit MIPS IV ISA that adds  new instructions allowing byte-granularity enforcement of memory accesses. A memory capability is unforgeable and grants access to a certain memory range. CHERI can support a number of protection models, such as pointer safety~\cite{Woodruff14} and software compartmentalization~\cite{Watson15,Tsampas17}. At time of writing, CHERI has only been realized as a soft microprocessor prototype on a 64-bit MIPS FPGA\@.
\emph{Hardware-Assisted Data-flow Isolation} (HDFI)~\cite{Song16} is a tagged memory extension for the RISC-V instruction set architecture that provides instruction-level granularity isolation and the ability to enforce a variety of security models (including pointer integrity). HDFI is efficient (< 2\% overhead) but only supports two simultaneous protection domains.
\fi

Only a few commercial processors, such as the SPARC M7\footnote{\url{https://swisdev.oracle.com/_files/What-Is-ADI.html}}, support tagged memory, which can be used to realize variety of security models (including pointer integrity). ARM recently announced support for memory tagging in the ARMv8.5-A architecture\footnote{\url{https://community.arm.com/processors/b/blog/posts/arm-a-profile-architecture-2018-developments-armv85a}}. It enforces that all accesses to memory must be made via a pointer with the correct tag.
Pointer tags use the existing address tagging feature in the ARM ISA that partly overlaps with the bits used to store PA PACs, meaning that enabling both features simultaneously reduces the available PAC size by eight bits.

Hardware-assisted memory tagging is designed primarily as a statistical debug aid against use-after-free and other temporal memory errors. \emph{Hardware-Assisted AddressSanitizer} (HWASAN)~\cite{Serebryany18} is an AArch64-specific compiler-based tool that builds upon \emph{AddressSanitizer} (ASAN) --- a memory-error detector popular for vetting memory safety bugs during software testing. ASAN can detect both spatial and temporal memory errors. HWASAN can leverage hardware tagged memory, such as SPARC ADI and the upcoming ARMv8.5-A to reduce the performance overhead associated with managing tagged memory checks in software. ASAN / HWASAN are complementary to \SHORTNAME, as they provide spatial and temporal safety for data accesses via pointers.

\emph{Intel Memory Protection Extensions} (MPX) is a hardware feature for detecting spatial memory errors that debuted in the Intel Skylake microarchitecture.
MPX is similar to the software based SoftBound~\cite{Nagarakatte09} and its hardware-based predecessor~\cite{Devietti08}.
Although Intel MPX is a hardware-assisted approach specifically designed to provide spatial memory safety guarantees, it is not faster than software-based approaches~\cite{Oleksenko17}.
It can cause up to 4x slowdown in the worst case with an average run-time overhead of 50\%.
It also suffers from other shortcomings, such as the lack of support for multithreading and several common C/C++ idioms.
GCC has dropped support for MPX altogether\footnote{\url{https://gcc.gnu.org/viewcvs/gcc?view=revision&revision=261304}}.

\paragraph*{Control-flow integrity.}

Carlini et al.~\cite{Carlini15} define \emph{fully-precise static CFI} as follows: \textit{``An indirect control-flow transfer along some edge is allowed only if there exists a non-malicious trace that follows that edge.''}
In other words, fully-precise static CFI enforces that execution follows a CFG that contains an edge if and only if that edge is exercised by intended program behavior.
Fully-precise static CFI is thus the most restrictive \emph{stateless} policy possible without breaking intended functionality.
To date, there exist no implementation of fully-precise CFI\@; all practical implementations are limited by the precision of CFGs obtained through static \emph{control analysis}.

Carlini et al.\ further show that all stateless CFI schemes, including fully-precise static CFI are vulnerable to \emph{control-flow bending}; attacks where each control-flow transfer is within a valid CFG, but where the program execution trace conforms to no feasible benign execution trace. For instance, in a stateless policy such as fully-precise static CFI, the best possible policy for return instructions (i.e., backward edges in the CFG) is to allow return instructions within a function $F$ to target any instruction that follows a call to $F$. In other words, fully-precise static CFI checks if a given control-flow transfer conforms to any of the known control-flow transfers from the current position in the CFG, and does not distinguish between different paths in the CFG that lead to a given control-flow transfer.
\ifabridged{}
For this reason CFI is typically augmented with a \emph{shadow call stack}~\cite{Abadi09,Davi12} to enforce integrity of return addresses stored on the call stack. We compare \SHORTNAME{} to CFI solutions in Section~\ref{sec:delta-to-fully-precise-cfi}.
\else

The seminal work on \emph{stateful CFI}~\cite{Abadi09} combines the restriction of indirect call instructions to valid targets within the CFG with a \emph{shadow call stack} to enforce integrity of return addresses stored on the call stack. The shadow stack maintains a shadow copy of each return address on the call stack in a separate region of memory the attacker cannot access. Each return instruction is then instrumented to validate that the returns addresses on the call and shadow stack match. This ensures that each return only returns to its corresponding call site.

\emph{Context-sensitive CFI}~\cite{vanderVeen15,Ding17}\label{sec:backward-edge-cfi}
is a generalization of stateful CFI techniques. It provides stronger security properties than stateless CFI\@.
For instance, \emph{path-sensitive} CFI~\cite{Ding17} can ensure that each control-flow transfer taken by the program is consistent with a non-malicious trace.
Context-sensitive CFI does not rely on data integrity, and can thus be enforced more efficiently than full data-integrity.
Nevertheless, context-sensitive has been dismissed as impractical for real-world adoption~\cite{Abadi09}.
Recent context-sensitive CFI implementations~\cite{vanderVeen15,Ding17} that rely on branch recording features available in modern 64-bit Intel microprocessors show promise in enabling context-sensitive CFI enforcement with reasonable overhead on commodity hardware. However, state-of-the-art implementations are either limited in terms of the size of of the branch history used to make CFI decisions, over-approximation of program CFG, or reliance on complex run-time monitoring, none of which are likely to be acceptable for integration into commodity operating systems.
\emph{uCFI} is a recent context-sensitive CFI scheme that uses hardware-assisted tracing capabilities on Intel processors to achieves precise CFI with low overhead by combining compile-time analysis with optimized run-time tracing, but is still reliant on a separate monitoring process~\cite{Hu18}.
Similarly to all CFI solutions, uCFI and context-sensitive CFI cannot protect against non-control data attacks that do not influence the program's execution trace.

\paragraph*{Control-flow integrity on ARM.}

\emph{CFI for Clang} provides a number of CFI schemes for C and C++ using LLVM and its Clang compiler fronted.
\footnote{\url{https://clang.llvm.org/docs/ControlFlowIntegrity.html}}.
For C code, it checks that function calls target a function of the correct type and uses a shadow stack to protect backward edges.
\emph{MoCFI}~\cite{Davi12} is a software-based CFI approach targeting ARM-based smartphone platforms; it uses a combination of a shadows stack, static analysis, and run-time heuristics to determine the set of valid targets for control-flow transfers.
Specifically, indirect functions calls are constrained to target instructions at the beginning of functions (as determined by static analysis) and indirect jumps (e.g.,into branch tables) are restricted to the function's scope.
However, MoCFI makes no attempt to protect the integrity of the shadow stack data, and is thus susceptible to data-oriented attacks that can break shadow stack integrity.
\emph{CFI CaRE}~\cite{Nyman17a} is a CFI solution targeting small, embedded ARM-based microcontrollers (MCUs).
Similarly to MoCFI, it uses a shadow stack, but accommodates small MCUs by relaxing the restriction on indirect calls to only validate that each call targets the beginning of functions.
In contrast to MoCFI, CFI CaRE uses the ability to perform hardware-enforced isolated execution on ARM MCUs to isolate the shadow stack from the protected program.

\fi

\section{Comparison with other integrity policies}
\label{sec:discussion}

\subsection{Fully precise pointer integrity}
\label{sec:fully-precise-pointer-integrity}

As discussed in Section~\ref{sec:pointer-integrity}, Pointer Integrity can be loosely defined as a policy ensuring that the value of a pointer at the time of use (dereference or call) corresponds to the value of the pointer when it was created. In this section, we provide a more rigorous definition of Pointer Integrity.

We define \emph{fully-precise pointer integrity} as follows: A pointer dereference is allowed if and only if the pointer is based on its target object.
We adopt Kuznetsov et al.'s~\cite{Kuznetsov14} definition of ``\emph{based on}'' and say a pointer $P$ \emph{is based} on a target object $X$ if, and only if, $P$ is obtained at run-time by \textit{''(i) allocating $X$ on the heap, (ii) explicitly taking the address of $X$, if $X$ is allocated statically, such as a local or global variable, or is a control-flow target (including return locations, whose addresses are implicitly taken and stored on the stack when calling a function), (iii) taking the address of a sub-object $y$ of $X$ (e.g., a field in the struct $X$), or (iv) computing a pointer expression (e.g., pointer arithmetic, array indexing, or simply copying a pointer) involving operands that are either themselves based on object $X$ or are not pointers.``}

Kuznetsov et al's CPI~\cite{Kuznetsov14} (Section~\ref{sec:related-work}) provides fully precise integrity guarantees for \emph{code pointers} by ensuring that accesses to sensitive pointers are safe (sensitive pointers are code pointers and pointers that may later be used to access sensitive pointers). However, CPI requires dedicated, integrity-protected storage for sensitive pointers.

As discussed in Section~\ref{sec:sec-eval}, \SHORTNAME{}, and \ARMPA{} solutions in general, achieve an approximation of fully-precise pointer integrity.
In particular, \SHORTNAME{} allows the substitution of a pointer $P$ by another pointer $P'$ \emph{based on} object $X$, if $P$ and $P'$ share the PA modifier.
In other words, when PA modifiers are unique to each protected pointer value, PA provides fully-precise pointer integrity. However, ensuring the uniqueness of PA modifiers is not possible in practice due to the following reasons: 
\begin{inparaenum}[1)]
\item program semantics may require a set of pointers to be substitutable with each other (e.g., pointers to callback functions)
\item the choice of allowed pointers may depend on run-time properties (e.g., which callback function was registered earlier).
\end{inparaenum}
In these cases, a unique modifier must be determined at run-time. Fully-precise pointer integrity does \emph{not} imply \emph{memory safety}. In the case of \PA, if the modifier is determined at run-time and stored in memory, the PA modifier itself may become a target for an attacker wishing to undermine the integrity policy. To avoid this, modifier values must be derived in a way which leaves the value outside the control of the attacker, e.g., stored in a dedicated hardware register, or read-only program memory.

\subsection{Fully-precise static CFI}
\label{sec:delta-to-fully-precise-cfi}

In contrast to stateless CFI, which allows control-flow transitions present in its CFG regardless of the origin of the code pointer value, PA-based solutions (including \SHORTNAME) can preclude forged pointer values from outside the process.
The policy that prevents pointer reuse can suffer from limitations similar to those present stateless CFI\@.

\SHORTNAME{} return address signing provides strong guarantees even when subjected to pointer reuse.
In contrast, a stateless CFI policy allows a function to return to any of its call sites.
As such, static CFI cannot prevent injection of pointers that are within the expected CFG, i.e., control-flow bending attacks.
\SHORTNAME{} additionally requires matching SP values, and that the reused return address originates from a prior function invocation of the same function within the same process for an attack to succeed.

\SHORTNAME{} forward-edge code pointer integrity provides similar guarantees (under reuse attacks) as LLVM's type-based protection (when subjected to any forged pointer).
In both cases, attacks are limited to using pointers of the correct dynamic type.
\SHORTNAME{} in addition requires that the injected pointer originates from the victim process.

\ifnotabridged{}
Path-sensitive CFI (Section~\ref{sec:backward-edge-cfi}) can provide stronger policies compared to both stateless CFI and PA-based solutions but current implementations use either extensive run-time monitoring or a shadow stack. In order for a shadow stack to be effective, it must be protected from modification by the attacker. This can be achieved by software instrumentation that sanitizes all memory accesses, hardware support for per-instruction memory isolation, or randomization.
\fi
While shadow-stacks protected through randomization can be implemented with minimal performance overhead, our adversary model precludes this approach.
Furthermore, software-isolated shadow stack solutions impose impractical performance overheads, and ARM processors do not currently provide direct hardware support for shadow stacks.

\section{Conclusion and Future Work}
\label{sec:future-work}

We plan to extend \SHORTNAME{} protection architecture to other protection domains like the OS kernel, or hypervisor.
\ifnotabridged{}
Such additions require that key configuration is trapped in on a higher exception level (EL), i.e., in the hypervisor or trusted software.
Trapping key configuration beyond the kernels reach prevents the kernel from updating \PA{} keys, and thus, from handling context switches.
Nonetheless, the
\else
The
\fi
only significant change for \SHORTNAME{} architecture is to arrange for key configuration for both kernel and EL0 \SHORTNAME{} to be trapped (and managed) on a higher exception level (EL2,3).
We are further looking at adding C++ support \SHORTNAME{}.
While we do not expect any fundamental problems, some C++ specific features, such as inheritance, cannot be directly handled by our current instrumentation strategy.
Authenticated pointers with PACs cannot be used by legacy code (Section~\ref{sec:pa}) while \SHORTNAME{}-instrumented code will trap if pointers without PACs are used.
For legacy and \SHORTNAME{} code to interact, we can use wrappers that manipulate function arguments and return values by embedding/stripping PACs.
For shared pointers or complex data structures, annotations can disable authentication of selected pointers, allowing programmers to manually adjust pointer conversion to and from legacy code.

Currently, the \SHORTNAME{} compiler assumes shared libraries to be uninstrumented.
Instrumented shared libraries must deal with PACs for statically allocated pointers after linking, and thus require changes to the dynamic linker.
\ifnotabridged{}
Moreover, if a future \PA-scheme utilizes dynamic modifiers for shared objects, the dynamic linker could then harmonize \PA{} modifiers among all callees using the shared resources.
\fi

Pointer integrity does not imply full memory safety (Section~\ref{sec:fully-precise-pointer-integrity}). Although \ARMPAFULL{} does not support bounds checking for pointer accesses with authenticated pointers, it has a general-purpose instruction, \instr{pacga}, for producing and validating PACs computed over the contents of two 64-bit registers.
This can be used to build authenticated canaries to identify buffer overflow attacks, or to validate the integrity (freshness) of atomic data, such as integer or counter
values.
In principle, \instr{pacga} instructions can even be chained to validate arbitrary-sized blocks of data.

Finally, effective ways of complementing \PA{} with other emerging memory safety mechanisms like the forthcoming support for memory tagging in ARMv8.5-A is an important line of future work.

\ifnotanonymous
\section*{Acknowledgments}

This work was supported in part by the Academy of Finland under grant nr. 309994 (SELIoT), and the Intel Collaborative Research Institute for Collaborative Autonomous \& Resilient Systems (ICRI-CARS).

The authors thank Kostya Serebryany and Rémi Denis-Courmont for interesting discussions and Zaheer Gauhar for implementation assistance.

\fi

{\normalsize\bibliographystyle{acm}
\bibliography{references}}
\appendix

\section{nbench experimental setup}
\label{sec:appendix-nbench}

The nbench benchmarks employs dynamic workload adjustment to allow the tests to expand or contract depending on the capabilities of the system under test. To achieve this, nbench employs timestamping to ensure that a test run exceeds a pre-determined minimum execution time. If a test run finishes before the minimum execution time has been reached, the test dynamically adjusts its workload, and tries again. For example, the Numeric Sort test will construct an array filled with random numbers, measure the time taken to sort the array. If the time is less than the pre-determined minimum time, the test will build two arrays, and try again. If sorting two arrays takes less time than the pre-determined minimum, the process repeats with more arrays. 

Since we want to determine the relative overhead in execution time caused by our instrumentation, we employ the methodology described by Brasser et al.~\cite{Brasser17} and modify nbench to instead run each test a constant number of iterations. The number of iterations was determined individually for each test based on the iteration counts determined by a unmodified nbench run on the FVP. We then instrument the nbench benchmarks using our PA-analogue (Section~\ref{sec:eval-pa-analogue}) and measure the relative execution time between non-instrumented and instrumented nbench tests on the HiKey development platform using the BusyBox \texttt{time} utility.

Each individual benchmark test was run 200 times using the pre-determined number of iterations.
Figure~\ref{fig:nbench_results_a}, in Section~\ref{sec:eval-nbench} shows instrumentation overhead for individual tests in relation to the uninstrumented test run. Table~\ref{tab:nbench-ratio} shows the numeric overhead ratio for each individual test. Because the nbench benchmarks are designed to measure performance in a manner which is operating system agnostic, they are written in ANSI C and only execute in a single thread. We therefore only consider user time when measuring the overhead of the instrumentation, and exclude context switches and system calls.

The run-time overhead of \SHORTNAME is dependent on specific run-time events, such as the number of function invocations in the case of return address signing. Figure~\ref{fig:nbench_results_b} in Section~\ref{sec:eval-nbench} shows the order of magnitude of instrumented run-time events in the nbench tests. We also report the  user mode run-time for uninstrumented nbench tests, the number of iterations of each individual test, and number of instrumented run-time events in Table~\ref{tab:nbench-stats}.

\section{SPEC CPU2017 experimental setup}
\label{sec:appendix-spec}

Due to unmanageable simulation times in the FVP simulator we have verified the correctness of \SHORTNAME{} instrumentation only on a subset of SPEC CPU2017 benchmarks.
Specifically, we chose the 505.mcf\_r and 519.lbm\_r benchmarks from the SPECrate 2017 integer and floating point suites, because these were the smallest C benchmarks in terms of lines of code.
The benchmarks were compiled using SPEC \texttt{runcpu}, with a AArch64-specific configuration specifying whole-program-llvm\footnote{\url{https://github.com/travitch/whole-program-llvm}}, with our \SHORTNAME{}-enabled LLVM, as the compiler.
We then extracted the bitcode --- created by whole-program-llvm during compilation --- and used it to instrument and compile the binaries we used for evaluation: one uninstrumented, one instrumented with PA instructions, and one instrumented with our PA-analogue.
We enabled both return address and forward-edge code pointer signing for the instrumented binaries.

We run the \SHORTNAME{}-instrumented binaries on the FVP simulator to confirm correct functionality.
The simulation time for the tested benchmarks was between 12 and 48 hours.
Performance benchmarks, for baseline and PA-enabled binaries, were run on the HiKey devices, using the same setup as our nbench evaluation.
The results are shown in Table~\ref{tab:spec-results}, and are based on five runs of each benchmark. In 505.mcf\_r we observed overheads consistent with our results from nbench. We observed no discernible overhead in 519.lbm\_r. We attributed this to the following properties of 519.lbm\_r:  
\begin{inparaenum}[(a)]
\item it does not exhibit forward-edge code pointers, and 
\item it has few non-leaf function calls in relation to the arithmetic computation performed part of the benchmark. 
\end{inparaenum}

\begin{table}
\centering
\caption{Overhead as ratio and standard deviation ($\sigma$) for return address signing and (forward-edge) code pointer signing for 505.mcf\_r and 519.lbm\_r SPEC benchmarks.}
\label{tab:spec-results}

\resizebox{\columnwidth}{!}{
\begin{tabular}{l|c|c|c|c}

\toprule

Benchmark & \multicolumn{2}{c|}{Uninstrumented} & \multicolumn{2}{c}{ret. addr. sign. + code ptr. integrity} \\

& ratio & $\sigma$ & ratio & $\sigma$ \\

\midrule

505.mcf\_r & 1 & 0.004 & 1.005 & 0.004 \\
519.lbm\_r & 1 & 0.000 & 1.000 & 0.000 \\

\bottomrule

\end{tabular}}
\end{table}

\begin{table*}
\centering
\caption{Overhead as ratio and standard deviation ($\sigma$) for nbench tests reported separately for uninstrumented, return address signing, (forward-edge) code pointer signing, data pointer signing and all instrumentation enabled.}
\label{tab:nbench-ratio}
\begin{tabular}{l|c|c|c|c|c|c|c|c|c|c}

\toprule
  \multirow{3}{*}{Test} &
  \multicolumn{2}{c|}{\multirow{2}{*}{Uninstrumented}} &
  \multicolumn{8}{c}{\SHORTNAME} \\

  &
  \multicolumn{2}{c|}{} &
  \multicolumn{2}{c|}{ret. addr. sign} &
  \multicolumn{2}{c|}{code ptr. signing} &
  \multicolumn{2}{c|}{data ptr. signing} &
  \multicolumn{2}{c}{all enabled} \\

  & ratio & $\sigma$ &
  ratio & $\sigma$ &
  ratio & $\sigma$ &
  ratio & $\sigma$ &
  ratio & $\sigma$ \\

\midrule

Numeric sort     & 1 & 0.002 & 1     & 0.003 & 1     & 0.003 & 1.293 & 0.003 & 1.293 & 0.003 \\ 
String sort      & 1 & 0.002 & 1.01  & 0.002 & 1     & 0.002 & 1.251 & 0.002 & 1.259 & 0.002 \\ 
Bitfield         & 1 & 0.002 & 1     & 0.002 & 1     & 0.002 & 1.15  & 0.002 & 1.15  & 0.001 \\ 
FP emulation     & 1 & 0.001 & 1     & 0.001 & 1     & 0.001 & 1.395 & 0.001 & 1.396 & 0.001 \\ 
Fourier          & 1 & 0.002 & 1.027 & 0.004 & 0.999 & 0.003 & 0.998 & 0.002 & 1.016 & 0.003 \\ 
Assignment       & 1 & 0.001 & 1     & 0.002 & 1     & 0.002 & 1.145 & 0.002 & 1.145 & 0.002 \\ 
Idea             & 1 & 0.001 & 1.004 & 0.002 & 1     & 0.002 & 1.279 & 0.002 & 1.283 & 0.002 \\ 
Huffman          & 1 & 0.001 & 0.999 & 0.001 & 0.999 & 0.001 & 1.294 & 0.001 & 1.295 & 0.002 \\ 
Neural net       & 1 & 0.001 & 1.002 & 0.002 & 1     & 0.002 & 1.001 & 0.002 & 1.001 & 0.003 \\ 
Lu decomposition & 1 & 0.001 & 1     & 0.002 & 1     & 0.002 & 1.173 & 0.002 & 1.173 & 0.002 \\ 

\midrule

\textbf{Geometric average} & 1 & - & 1.004 & - & 1.000 & - & 1.191 & - & 1.195 & - \\

\bottomrule

\end{tabular}
\end{table*}

\begin{table*}
\centering
\caption{User mode run-time (utime) and standard deviation ($\sigma$) in seconds for uninstrumented nbench tests, the pre-determined number of iterations for each individual test, and the number of run-time events that are affected by instrumentation. Non-leaf calls correspond to function invocations protected by return address signing. Leaf calls correspond to function invocations which do no store the value of LR in memory, and thus can be left uninstrumented. Instruction pointers created and indirect calls are instrumented by (forward-edge) code pointer signing, and data pointer loads / stores correspond to events where data pointer instrumentation is active.}

\label{tab:nbench-stats}

\resizebox{\textwidth}{!}{
\begin{tabular}{l|c|c|r|r|r|c|c|r}

\toprule
  \multirow{2}{*}{Test} &
  \multicolumn{3}{c|}{Baseline} &
  \multicolumn{5}{c}{Instrumented events} \\

& utime & $\sigma$ & iterations &
non-leaf calls & leaf calls & instr. ptr. created & indirect calls & data ptr. ldr/str \\

\midrule

Numeric sort     & 3.573 & 0.007 &       350 &    1802 &  7117598 & 10 & 5  & 302212833 \\ 
String sort      & 2.971 & 0.005 &       125 & 3977237 &  1022510 & 10 & 5  & 180105579 \\ 
Bitfield         & 2.687 & 0.004 & 101647890 &    5669 &     4308 & 10 & 5  & 104670943 \\ 
FP emulation     & 5.862 & 0.004 &        35 &  616536 & 37906118 & 10 & 5  & 589518589 \\ 
Fourier          & 2.693 & 0.005 &     25870 & 5240188 &      161 & 10 & 5  &     27504 \\ 
Assignment       & 4.414 & 0.005 &        10 &  225602 &   113353 & 10 & 5  & 190662093 \\ 
Idea             & 2.808 & 0.004 &      1500 & 1640184 & 54420196 & 10 & 5  & 196844406 \\ 
Huffman          & 4.212 & 0.005 &      1000 &   17659 & 46983276 & 10 & 5  & 343176061 \\ 
Neural net       & 5.477 & 0.007 &        10 &  359423 &   441412 & 10 & 5  &       782 \\ 
Lu decomposition & 3.596 & 0.005 &       230 &   18970 &   441412 & 10 & 5  & 186704928 \\ 

\bottomrule

\end{tabular}}

\end{table*}

\begin{table*}
\section{\ARMPAFULL Instructions}
\label{sec:appendix-instructions}
\centering
  \caption{List of PA instructions \ifabridged referred to in the main paper\fi~\cite{ARMv8A}. \emph{PA Key} indicates the PA key the instruction uses. \emph{Addr.} indicates the source of the address to be signed / authenticated (\textit{Xd} indicates that the address is specified using a general purpose register). \emph{Mod.} indicates the modifier used by the instruction (\textit{Xm} indicates that the modifier is specified by a general purpose register.) The \emph{backwards-compatible} column indicates if the instruction encoding resides in the NOP space for pre-existing ARMv8-A processors.}
\label{tab:pa-instructions}

\resizebox{0.7\textwidth}{!}{
\begin{tabular}{c|l|c|c|c|c|c|c|c|c}
\toprule
  \multirow{3}{*}{Instruction} &
  \multirow{3}{*}{Mnemonic} &
  \multicolumn{5}{c|}{PA Key} &
  \multirow{3}{*}{Addr.} &
  \multirow{3}{*}{Mod.} &
  \multirow{3}{*}{\parbox{1.5cm}{Backwards-compatible}} \\
  & & \multicolumn{2}{c|}{Instr.} & \multicolumn{2}{c|}{Data} & Gen-& & & \\
  & & A & B & A & B & eric & & & \\
\midrule
  \multicolumn{10}{c}{BASIC POINTER AUTHENTICATION INSTRUCTIONS} \\
\midrule
\ifabridged
  \multirow{4}{*}{\parbox{3.5cm}{Add PAC to instr. addr.}}
\else
  \multirow{10}{*}{\parbox{3.5cm}{Add PAC to instr. addr.}}
\fi
  & \instr{paciasp}   & \tableYes{} & & & & & LR          & SP            & \tableYes{} \\
  & \instr{pacia}     & \tableYes{} & & & & & \textit{Xd} & \textit{Xm}   & \tableNo{} \\
\ifnotabridged
  & \instr{paciaz}    & \tableYes{} & & & & & LR          & \textit{zero} & \tableYes{} \\
  & \instr{paciza}    & \tableYes{} & & & & & \textit{Xd} & \textit{zero} & \tableNo{} \\
  & \instr{pacia1716} & \tableYes{} & & & & & X17         & X16           & \tableYes{} \\
\fi
  & \instr{pacibsp}   & & \tableYes{} & & & & LR          & SP            & \tableYes{} \\
  & \instr{pacib}     & & \tableYes{} & & & & \textit{Xd} & \textit{Xm}   & \tableNo{} \\
\ifnotabridged
  & \instr{pacibz}    & & \tableYes{} & & & & LR          & \textit{zero} & \tableYes{} \\
  & \instr{pacizb}    & & \tableYes{} & & & & \textit{Xd} & \textit{zero} & \tableNo{} \\
  & \instr{pacib1716} & & \tableYes{} & & & & X17         & X16           & \tableYes{} \\
\fi
\midrule
\ifabridged
  \multirow{2}{*}{\parbox{3.5cm}{Add PAC to data addr.}}
\else
  \multirow{4}{*}{\parbox{3.5cm}{Add PAC to data addr.}}
\fi
  & \instr{pacda}    & & & \tableYes{} & & & \textit{Xd} & \textit{Xm,}  & \tableNo{} \\
\ifnotabridged
  & \instr{pacdza}   & & & \tableYes{} & & & \textit{Xd} & \textit{zero} & \tableNo{} \\
\fi
  & \instr{pacdb}    & & & & \tableYes{} & & \textit{Xd} & \textit{Xm}   & \tableNo{} \\
\ifnotabridged
  & \instr{pacdzb}   & & & & \tableYes{} & & \textit{Xd} & \textit{zero} & \tableNo{} \\
\fi
\midrule
  \multirow{1}{*}{\parbox{3.5cm}{Calculate generic MAC}}
  & \instr{pacga} & & & & & \tableYes{} & & & \tableNo{} \\
\midrule
\ifabridged
  \multirow{4}{*}{\parbox{3.5cm}{Authenticate instr. addr.}}
\else
  \multirow{10}{*}{\parbox{3.5cm}{Authenticate instr. addr.}}
\fi
  & \instr{autiasp}   & \tableYes{} & & & & & LR          & SP            & \tableYes{} \\
  & \instr{autia}     & \tableYes{} & & & & & \textit{Xd} & \textit{Xm}   & \tableNo{} \\
\ifnotabridged
  & \instr{autiaz}    & \tableYes{} & & & & & LR          & \textit{zero} & \tableYes{} \\
  & \instr{autiza}    & \tableYes{} & & & & & \textit{Xd} & \textit{zero} & \tableNo{} \\
  & \instr{autia1716} & \tableYes{} & & & & & X17         & X16           & \tableYes{} \\
\fi
  & \instr{autibsp}   & & \tableYes{} & & & & LR          & SP            & \tableYes{} \\
  & \instr{autib}     & & \tableYes{} & & & & \textit{Xd} & \textit{Xm}   & \tableNo{} \\
\ifnotabridged
  & \instr{autibz}    & & \tableYes{} & & & & LR          & \textit{zero} & \tableYes{} \\
  & \instr{autizb}    & & \tableYes{} & & & & \textit{Xd} & \textit{zero} & \tableNo{} \\
  & \instr{autib1716} & & \tableYes{} & & & & X17         & X16           & \tableYes{} \\
\fi
\midrule
\ifabridged
  \multirow{2}{*}{\parbox{3.5cm}{Authenticate data addr.}}
\else
  \multirow{4}{*}{\parbox{3.5cm}{Authenticate data addr.}}
\fi
  & \instr{autda}    & & & \tableYes{} & & & \textit{Xd} & \textit{Xm,}  & \tableYes{} \\
\ifnotabridged
  & \instr{autdza}   & & & \tableYes{} & & & \textit{Xd} & \textit{zero} & \tableYes{} \\
\fi
  & \instr{autdb}    & & & & \tableYes{} & & \textit{Xd} & \textit{Xm}   & \tableYes{} \\
\ifnotabridged
  & \instr{autdzb}   & & & & \tableYes{} & & \textit{Xd} & \textit{zero} & \tableYes{} \\
\midrule
  \multirow{3}{*}{\parbox{3.5cm}{Strip PAC}}
  & \instr{xpacd}   & & & & & & \textit{Xd} & & \tableNo{} \\
  & \instr{xpaci}   & & & & & & \textit{Xd} & & \tableNo{} \\
  & \instr{xpaclri} & & & & & & LR          & & \tableYes{} \\
\fi
\midrule
  \multicolumn{10}{c}{COMBINED POINTER AUTHENTICATION INSTRUCTIONS} \\
\midrule
  \multirow{2}{*}{\parbox{3.5cm}{Authenticate instr. addr. and return}} &
    \instr{retaa}    & \tableYes{} & & & & & LR & SP & \tableNo{} \\
  & \instr{retab}    & & \tableYes{} & & & & LR & SP & \tableNo{} \\
\midrule
\ifabridged
  \multirow{2}{*}{\parbox{3.5cm}{Authenticate instr. addr. and branch}}
\else
  \multirow{4}{*}{\parbox{3.5cm}{Authenticate instr. addr. and branch}}
\fi
  & \instr{braa}     & \tableYes{} & & & & & \textit{Xd} & \textit{Xm}   & \tableNo{} \\
\ifnotabridged
  & \instr{braaz}    & \tableYes{} & & & & & \textit{Xd} & \textit{zero} & \tableNo{} \\
\fi
  & \instr{brab}     & & \tableYes{} & & & & \textit{Xd} & \textit{Xm}   & \tableNo{} \\
\ifnotabridged
  & \instr{brabz}    & & \tableYes{} & & & & \textit{Xd} & \textit{zero} & \tableNo{} \\
\fi
\midrule
\ifabridged
  \multirow{2}{*}{\parbox{3.5cm}{Authenticate instr. addr. and branch with link}}
\else
  \multirow{4}{*}{\parbox{3.5cm}{Authenticate instr. addr. and branch with link}}
\fi
  & \instr{blraa}    & \tableYes{} & & & & & \textit{Xd} & \textit{Xm}   & \tableNo{} \\
\ifnotabridged
  & \instr{blraaz}   & \tableYes{} & & & & & \textit{Xd} & \textit{zero} & \tableNo{} \\
\fi
  & \instr{blrab}    & & \tableYes{} & & & & \textit{Xd} & \textit{Xm}   & \tableNo{} \\
\ifnotabridged
  & \instr{blrabz}   & & \tableYes{} & & & & \textit{Xd} & \textit{zero} & \tableNo{} \\
\fi
\midrule
  \multirow{2}{*}{\parbox{3.5cm}{Authenticate instr. addr. and exception return}}
  & \instr{eretaa}   & \tableYes{} & & & & & ELR & SP & \tableNo{} \\
  & \instr{eretab}   & & \tableYes{} & & & & ELR & SP & \tableNo{} \\
\midrule
  \multirow{2}{*}{\parbox{3.5cm}{Authenticate data. addr. and load register}}
  & \instr{ldraa}    &  & & & \tableYes{} & & \textit{Xd} & \textit{zero} & \tableNo{} \\
  & \instr{ldrab}    & &  & & & \tableYes{} & \textit{Xd} & \textit{zero} & \tableNo{} \\
\bottomrule
\end{tabular}
}
\end{table*}

\end{document}